\documentclass[a4paper,twocolumn,11pt,accepted=2025-09-06]{quantumarticle}
\pdfoutput=1
\usepackage[utf8]{inputenc}
\usepackage[T1]{fontenc}
\usepackage[english]{babel}

\usepackage{amsmath,amsfonts,amssymb}
\usepackage{graphicx}
\usepackage[colorlinks=true, allcolors=blue]{hyperref}
\usepackage{braket}
\usepackage{mathtools}
\usepackage[dvipsnames]{xcolor}
\usepackage{tcolorbox}
\usepackage{enumerate}
\usepackage[framemethod=TikZ,xcolor=RGB]{mdframed}
\usepackage{bm}
\usepackage{multirow}

\definecolor{mybrown}{RGB}{102,101,71}
\definecolor{myyellow}{RGB}{255,226,138}
\definecolor{mygreen}{RGB}{111,203,159}

\hypersetup{
    colorlinks=true,
    linkcolor=blue,
    filecolor=magenta,      
    urlcolor=cyan,
    pdfpagemode=FullScreen,
    }

\newtheorem{theorem}{Theorem}
\newtheorem{corollary}{Corollary}
\newtheorem{definition}{Definition}[section]

\renewcommand{\vec}{\boldsymbol}
\newcommand{\de}[1]{\left(#1\right)}
\newcommand{\De}[1]{\left[#1\right]}

\newcommand{\nospacemod}{\hspace{-7pt}\mod}

\usepackage{tikzpagenodes}

\begin{document}

\title{Backpropagation scaling in parameterised quantum circuits}
%Evaluating gradients of quantum circuits in parallel
%Parameterised quantum circuits with fast gradient estimation

\author{Joseph Bowles}\thanks{joseph@xanadu.ai}
\author{David Wierichs}
\author{Chae-Yeun Park}

% \affil{Xanadu, Toronto, ON, M5G 2C8, Canada} % for article
\affiliation{Xanadu, Toronto, ON, M5G 2C8, Canada} % for revtex

% \date{}

\begin{abstract}
    The discovery of the backpropagation algorithm ranks among one of the most important moments in the history of machine learning, and has made possible the training of large-scale neural networks through its ability to compute gradients at roughly the same computational cost as model evaluation. Despite its importance, a similar backpropagation-like scaling for gradient evaluation of parameterised quantum circuits has remained elusive. Currently, the most popular method requires sampling from a number of circuits that scales with the number of circuit parameters,
    making training of large-scale quantum circuits prohibitively expensive in practice. Here we address this problem by introducing a class of structured circuits that are not known to be classically simulable and admit gradient estimation with significantly fewer circuits. 
    In the simplest case---for which the parameters feed into commuting quantum gates---these circuits allow for fast estimation of the gradient, higher order partial derivatives and the Fisher information matrix. Moreover, specific families of parameterised circuits exist for which the scaling of gradient estimation is in line with classical backpropagation, and can thus be trained at scale.
    In a toy classification problem on 16 qubits, such circuits show competitive performance with other methods, while reducing the training cost by about two orders of magnitude.

\end{abstract}

\maketitle

\section{Introduction}
Many tasks in optimisation and machine learning involve the minimisation of a continuous cost function over a set of parameters, and gradient based approaches are often the methods of choice to tackle such tasks. The efficiency with which gradients can be evaluated is crucial in these methods, since it sets a practical upper limit on the number of optimisation steps, determined by the time and cost of running the algorithm. The success of neural network models in machine learning for example is frequently credited to the possibility of fast gradient evaluation via backpropagation \cite{backprophinton,autodiff}, enabling the models to scale to the enormous sizes we see today. 

Within quantum computation, much hope has been placed on parameterised quantum circuits to provide a new generation of powerful ans\"atze for optimisation and machine learning tasks \cite{cerezo2021variational,benedetti2019parameterized}. As such, a wide array of variational quantum algorithms have been designed that hinge on the ability to estimate gradients of quantum circuits. Despite this, classes of circuits for which fast gradient estimation is possible are still missing. Currently, the standard method to compute gradients is via the parameter-shift method \cite{PS1,PS2,PS3,PS4}, which in the simplest case requires sampling from a pair of distinct circuits for every parameter in the model. For this reason, the cost of estimating the gradient can be significantly more expensive than evaluating the function itself. This is in stark contrast to automatic differentiation methods such as backpropagation \cite{autodiff}, which can allow for exact gradient computation with roughly the same time and memory required to evaluate the function itself.

Furthermore, the stochasticity of quantum models results in an additional sampling cost that scales inversely with the square of the precision of the gradient estimate \cite{PS2}. For problems with a large number of parameters, this quickly makes gradient estimation prohibitively expensive in practice. To give some concrete numbers, to perform one epoch of gradient descent for a dataset with 1000 data points using a quantum model with 1000 parameters to a gradient precision of $10^{-4}$ would require $2\times 10^{14}$ quantum circuit shots. With a quantum computer operating continually at clock speed of 1MHz, this would take around six years.

All of this suggests that variational quantum algorithms--and in particular quantum machine learning algorithms--are in dire need of ans\"atze that allow for faster gradient estimation than with the parameter-shift rule, and a number of works have explored other methods \cite{anastasiou2023really,basheer2022alternating,kasture2022protocols,frugal1,frugal2,frugal3,frugal4,ding2024random} and frameworks \cite{verdon2018universal} to this end. For generic, unstructured parameterised quantum circuits, fast gradient estimation appears difficult to achieve however, since backpropagation-like scaling is known to be impossible given access to only single copies of an unknown input state \cite{abbas2023quantum}. In order to make progress in this direction, it therefore seems necessary to focus attention on specific circuit structures. In some sense this should not be surprising: building successful machine learning models has always been about finding well-designed structures that are tailored for fast optimisation, and one should not expect this to change when moving to quantum models.   

In this work we present classes of parameterised quantum circuits that allow for faster gradient estimation than parameter-shift methods. In the simplest case (Sec.\ \ref{sec:commuting}), these circuits feature mutually commuting parameterised gates, which we call \emph{commuting-generator circuits}. Although gates in these circuits commute and can thus be simultaneously diagonalised, both the input state and final measurement need not be in this basis, which prevents classical simulability. The restriction to commuting gates allows the gradient estimation to be parallelised over the parameters of the circuit: whereas the parameter-shift method requires sampling from a pair of circuits for each parameter, our method returns the complete gradient vector to the same precision by sampling from a single circuit of the same number of qubits. This can lead to enormous savings in the number of circuit shots and in certain cases matches the scaling of backpropagation\footnote{We note that our work does not represent a quantum version of the classical backpropogation algorithm, but is a different approach with comparable scaling guarantees.}. Surprisingly, this class of circuits even allows for parallelisation when estimating the Fisher information matrix, and all partial derivatives of a given degree (at the cost of exponentially increasing variance with degree). This is something which even neural network models cannot offer, and opens up the possibility for efficient higher-order optimisation methods. 

We also show (Sec.\ \ref{sec:block}) how one can go beyond the restriction to commuting gates and design classes of circuits with non-commuting parameterised gates that also allow for parallel gradient estimation, which we call \emph{commuting-block circuits}. Here, the circuits must have a specific block structure, such that gates commute within blocks, with generators from distinct blocks having a fixed commuting or anticommuting relation. Due to this extra freedom, parameterised circuits from this class can be more expressive than their commuting counterparts. This raises the question of what are the ultimate limits of expressibility of these models, which we leave as an open question.  

We then present two specific examples of commuting-generator circuits and derive the corresponding circuits needed for the evaluation of partial derivatives (Sec.\ \ref{sec:examples}). The first of these circuits is used as a basis to construct a quantum model whose layers are equivariant with respect to translation permutations of qubits. This model is compared against another translation-equivariant model that features non-commuting gates, and against a quantum convolutional neural network at a classification task with translation symmetry (Sec.\ \ref{sec:numerics}). Simulating 16-qubit models, we show that the circuit with commuting generators achieves the lowest training cost and the best test accuracy of the three models, using orders of magnitude fewer circuit shots. 

Going forward, we hope our work encourages a move away from viewing generic unstructured quantum circuits as viable ans\"atze for optimisation and machine learning, and highlights both the need for and potential of creating quantum circuit ans\"atze with well-designed and fit-for-purpose structures. 

\begin{figure*}[t!]
    \centering
\includegraphics[width=\textwidth]{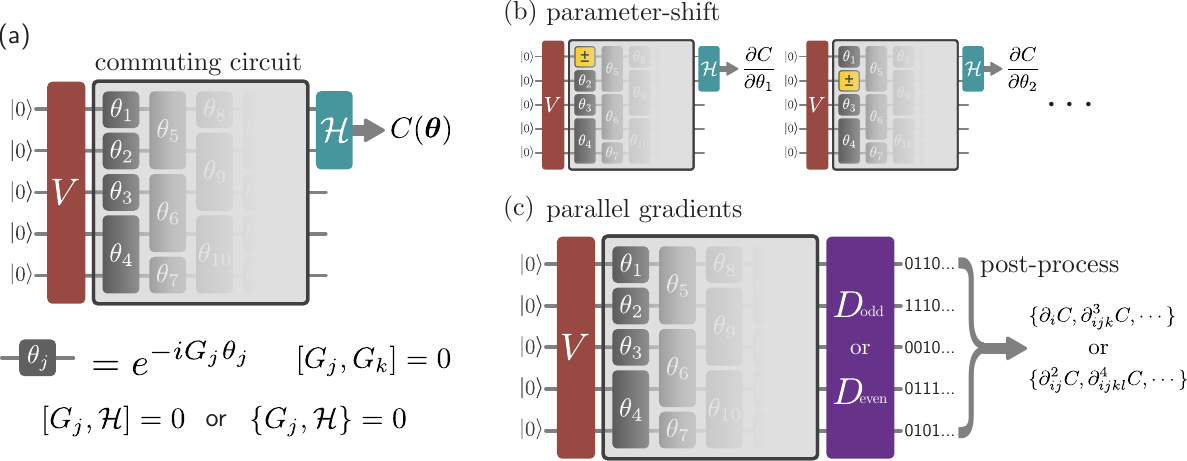}
    \caption{(a) The simplest type of circuit class we consider, which we call \emph{commuting-generator circuits}. An arbitrary unitary $V$ is applied, followed by a parameterised quantum circuit comprised of gates $\exp(-i\theta_j G_j)$ with mutually commuting generators $G_j$. An observable $\mathcal{H}$ is measured on some subset of the output qubits, and for each $G_j$ either (i) $\mathcal{H}$ commutes with $G_j$ or (ii) $\mathcal{H}$ anticommutes with $G_j$. (b) The standard method to estimate the gradient of quantum circuits is via parameter-shift rules. For gate generators with two distinct eigenvalues, the partial derivative with respect to each parameter is evaluated by estimating the difference between two circuits, where the parameter in question has been shifted. The resources required using this method can therefore be much larger than those required to estimate $C(\vec{\theta})$, which can be prohibitively expensive for circuits with thousands of parameters. (c) The corresponding circuit used for parallel gradient estimation. A unitary $D_{\text{odd}}$ (or $D_{\text{even}}$) is applied to the original circuit and the circuit is sampled from in the computational basis $M$ times. All odd (or even) partial derivatives can then be estimated to additive error $\epsilon = \mathcal{O}(\frac{1}{\sqrt{M}})$ by classically post processing the outcomes. In the case that all generators and the observable are stabiliser operators, the unitaries $D_{\text{even}}$, $D_{\text{odd}}$ are Clifford unitaries and the post-processing consists of evaluating expectation values of subsets of qubits only.}
    \label{fig:mainfig}
\end{figure*}

\section{Backpropagation scaling in parameterised quantum circuits}
The class of circuits we consider in this work take the form of classically parameterised quantum circuits:
\begin{align}\label{pqc}
    C(\vec{\theta}) = \bra{0}U^\dagger(\vec{\theta})\mathcal{H}U(\vec{\theta})\ket{0}
\end{align}
where 
\begin{align}
U(\vec{\theta})=\prod_j e^{-i\theta_jG_j}
\end{align}
is a unitary operator with Hermitian generators $G_j$ and classical parameters $\vec{\theta}$, and $\mathcal{H}$ is some Hermitian observable. On hardware, by measuring $\mathcal{H}$ and collecting $M$ shots, one obtains an unbiased estimate of $C(\vec{\theta})$ with variance scaling as $\mathcal{O}(1/M)$. 

To compute the gradient $\nabla C(\vec{\theta})$ of such circuits, the most common approach is to use \emph{parameter-shift methods} \cite{PS0, PS1, PS2, PS3, PS4, PS5}. In the simplest case of gate generators with two distinct eigenvalues, an unbiased estimate of the gradient can be obtained by evaluating two circuits for every parameter in the model, so that 
\begin{align}
\frac{\partial C}{\partial \theta_j}\propto C(\vec{\theta}+\vec{\alpha}_j)-C(\vec{\theta}-\vec{\alpha}_j),
\end{align}
with $\vec{\alpha_j}$ a vector with zeros on all components except the $j^{\text{th}}$. To obtain an estimate of the gradient to the same precision as $C(\vec{\theta})$, one therefore needs to take $M$ shots from each circuit, so that each component has variance  $\mathcal{O}(1/M)$. The cost of gradient estimation via this method is therefore larger than the cost of estimating $C(\vec{\theta})$ by a factor proportional to the number of trainable parameters. 

In similar spirit to \cite{abbas2023quantum}, we define a method to have \emph{backpropagation scaling} if it returns an unbiased estimate of $\nabla C(\vec{\theta})$ to the same precision as $C(\vec{\theta})$, 
with a logarithmic overhead in the number of parameters compared to the evaluation of $C(\vec{\theta})$.

\begin{definition}[backpropagation scaling]
    Consider a parameterised quantum circuit of the form \eqref{pqc} with $n$ parameters, which returns an unbiased estimate of $C(\vec{\theta})$ with variance $\mathcal{O}(\frac{1}{M})$ by sampling $M$ shots from the circuit. Denote by $\text{TIME}(C)$ and $\text{MEM}(C)$ the time and space complexity of this procedure, and by $\text{TIME}(\nabla C)$ and $\text{MEM}(\nabla C)$ the time and space complexity of obtaining an unbiased estimate of the gradient with elementwise variance $\mathcal{O}(1/M)$.
    Then we say that a gradient method has backpropagation scaling if 
    \begin{align}
        \text{TIME}(\nabla C)\leq c_t \text{TIME}(C)
    \end{align}
    and 
    \begin{align}
        \text{MEM}(\nabla C)\leq c_m \text{MEM}(C)
    \end{align}
    with $c_t,c_m= \mathcal{O}(\log(n))$.
\end{definition}

In this work, we will present families of circuits and gradient methods that achieve such scaling, even obtaining examples where $c_t,c_m$ have constant scaling. 

\section{Commuting-generator circuits}\label{sec:commuting}
The first class of circuits we consider consists of arbitrary input state preparation, mutually commuting parameterised gates, and measurement of an observable (see Fig.\ \ref{fig:mainfig}(a)). These circuits are thus given by 
\begin{align}\label{ansatz}
    C(\vec{\theta}) = \bra{0}V^\dagger U^{\dagger}(\vec{\theta})\mathcal{H}U(\vec{\theta}) V\ket{0},
\end{align}
where $V$ is an arbitrary unitary, $\mathcal{H}$ is an observable, and the parameterised unitary takes the form 
\begin{align}
    U(\vec{\theta})=\prod_{j=1}^n \exp(-iG_j\theta_j)
\end{align}
such that $[G_j,G_k]=0\;\forall\;j,k\in[n]$, where we adopt the notation $[n] = \{1,\cdots,n\}$. We further restrict our attention to observables $\mathcal{H}$ such that for each $G_j$ we have either $[\mathcal{H},G_j]=0$ or $\{\mathcal{H},G_j\}=0$; i.e. a given $G_j$ either commutes or anticommutes with $\mathcal{H}$. This is the case, for example, if all generators and observables are tensor products of Pauli operators. We remark that many applications involve evaluating sums of observables of this form, since any Hermitian multi-qubit operator can be decomposed in the Pauli basis. Here we will consider just a single $\mathcal{H}$ for simplicity however; evaluating gradients for sums can be achieved in the usual way by summing the contributions of individual observables. 

We will refer to such circuits as \emph{commuting-generator circuits}. Note that this class differs from the `commuting quantum circuits' \cite{bremner2016average,ni2012commuting} commonly studied in complexity theory due to the arbitrary unitary $V$ at the start of the circuit. Interestingly, circuits in this class allow for parallel gradient estimation in the following sense. 
\begin{theorem}\label{thm:grad}
Consider a commuting-generator circuit $C(\vec{\theta})$ of the form \eqref{ansatz}. %such that (i) $[G_j,G_k]=0\;\forall\;j,k\in [n]$, and (ii) each $G_j$ either commutes or anticommutes with $\mathcal{H}$. \
Then an unbiased estimator of the gradient
\begin{align}
    \nabla C(\vec{\theta})=\left(\frac{\partial C}{\partial \theta_1},\frac{\partial C}{\partial \theta_2},\cdots,\frac{\partial C}{\partial \theta_n}\right)
\end{align}
can be obtained by classically post-processing a single circuit $C'$ with the same number of qubits as $C$.
As the measurement statistics of $C'$ are used to estimate all derivatives simultaneously, the variance of each derivative estimator scales as $\mathcal{O}(1/M)$, where $M$ is the total number of shots used to evaluate $C'$.
\end{theorem}
\emph{Proof}---Since the generators commute, we have 
\begin{align}\label{thegrad}
    \frac{\partial C}{\partial \theta_j}= i \bra{\psi_0} U^{\dagger}(\vec{\theta})[G_j,\mathcal{H}]U(\vec{\theta})\ket{\psi_0},
\end{align}
where $\ket{\psi_0}=V\ket{0}$. The gradient with respect to $\theta_j$ therefore corresponds to measuring the expectation value of the observable $O_j=i[G_j,\mathcal{H}]$ on $\ket{\psi_0}$, which vanishes if $G_j$ and $\mathcal{H}$ commute. It follows that we only need to consider those generators that anticommute with $\mathcal{H}$. Consider two such generators $G_j, G_k$. Using anticommutativity we have $i[G_j,\mathcal{H}] = 2iG_j\mathcal{H}$ and so
\begin{align}
    [O_j,O_k] &= [i[G_j,\mathcal{H}],i[G_k,\mathcal{H}]] \nonumber\\
    &=-4(G_j\mathcal{H}G_k\mathcal{H}-G_k\mathcal{H}G_j\mathcal{H}) \nonumber \\
    &=0.
\end{align}
The operators $O_j$ therefore mutually commute and can be simultaneously diagonalised. Denoting by $\{\ket{\psi_i}\}$ the basis in which the $O_j$ are diagonal, we have
\begin{align}\label{odecomp}
    O_j = \sum_{i} \lambda_i(O_j)\ket{\psi_i}\bra{\psi_i}
\end{align}
with $\{\lambda_i(O_j)\}$ the corresponding eigenvalues of $O_j$. To estimate the gradients in parallel, we use the same circuit as $C(\vec{\theta})$ but measure in the basis $\{\ket{\psi_i}\}$, i.e.\ we sample from the distribution 
\begin{align}
    P(i) = \vert\bra{\psi_i}U(\vec{\theta})V\ket{0}\vert^2.
\end{align}
From \eqref{odecomp} and \eqref{thegrad} we have 
\begin{align}\label{gradexp}
    \frac{\partial C}{\partial \theta_j} = \mathbb{E}_{P(i)}[\lambda_i(O_j)].
\end{align}
We estimate $\frac{\partial C}{\partial \theta_j}$ by sampling $M$ shots $\{i_1,\cdots,i_M\}$ from $P(i)$ and estimating the expectation value \eqref{gradexp} by
\begin{align}\label{postprocess}
    \frac{\partial C}{\partial \theta_j} \approx \frac{1}{M}\sum_{k=1}^M \lambda_{i_k}(O_j). 
\end{align}
From the central limit theorem, since the variance of the sample mean of a bounded random variable scales as $\mathcal{O}(1/M)$, we have proven the claim. $\blacksquare$

Although the basis $\{\ket{\psi_i}\}$ in which the $O_j$ are diagonal must exist, one may still encounter a difficulty in finding or implementing the unitary that rotates to this basis. In some cases this procedure may be simple however. For this reason we will often focus on circuits whose generators and observables are Pauli products (tensor products of operators in $\{\mathbb{I},X,Y,Z\}
$) and will call this class of circuits \emph{commuting-Pauli-generator circuits}. For these circuits, the operators $O_j$ are also Pauli products, and this implies that the diagonalizing unitary can be implemented efficiently.
\begin{corollary}
Consider an N-qubit commuting-Pauli-generator circuit. Then the gradient can be estimated in parallel by a circuit ${C}'$ with the same number of qubits as ${C}$ and a depth at most $O\left(\frac{N}{\log N}\right)$ more than ${C}$.
\end{corollary}
The above depth bound follows from \cite[Cor 1.1]{jiang2020optimal}; note that one can also reduce the depth further to $O(\log N)$ at the expense of additional auxiliary qubits. Here, the diagonalising unitary $D$ is Clifford, and partial derivatives therefore correspond to evaluating products of Pauli $Z$ operators on subsets of qubits. Efficient algorithms to simultaneously diagonalise sets of stabiliser operators also exist \cite{crawford2021efficient,yen2020measuring,gokhale2019minimizing}, and for sets of operators with a lot of structure, suitable circuits can be found by hand (as we will see in Sec.\ \ref{sec:xgen}). 
% For example, in \cite{crawford2021efficient} it is shown that any collection of $n$ commuting $N$-qubit stabiliser operators can be diagonalised via a Clifford circuit with $O(Nn/\log n)$ two-qubit gates. 

Although the above depth bound scales with $N$, for some choices of generators and observables, the diagonalising unitary has constant depth. In these cases, the gradient evaluation therefore comes at the same computational cost as model evaluation, thus allowing for backpropagation scaling. In Sec.\ \ref{sec:examples} we will see two examples of such circuits. 

\subsection{Higher order partial derivatives}
By repeating the same process as in the proof of Theorem \ref{thm:grad}, one finds that the partial derivatives of any order can also be estimated in parallel.

\begin{theorem}\label{thm:pds}
Consider a commuting-generator circuit $C(\vec{\theta})$ as in Theorem \ref{thm:grad} and a fixed integer $t$. Unbiased estimates of all $t^{\text{th}}$ order partial derivatives of $C$ can be obtained by classically post-processing a single circuit $C'$ with the same number of qubits as $C$. The variance of each derivative estimator scales as $\mathcal{O}(1/M)$, where $M$ is the total number of shots used to evaluate $C'$.
\end{theorem}
\emph{Proof}---We start by considering the case $t=2$. Differentiating \eqref{thegrad} again we find that the second-order partial derivatives are
\begin{align}
    \frac{\partial C^2}{\partial \theta_j\partial\theta_k} 
    %&= i\bra{\psi_0} U^{\dagger}(\vec{\theta})[G_k,O_j]U(\vec{\theta})\ket{\psi_0} \nonumber \\
    &=-4\bra{\psi_0} U^{\dagger}(\vec{\theta})G_jG_k\mathcal{H}U(\vec{\theta})\ket{\psi_0}
\end{align}
if both $G_j$ and $G_k$ anticommute with $\mathcal{H}$, and are zero otherwise. Note that in the non-zero  case one has $[G_jG_k\mathcal{H},G_lG_m\mathcal{H}]=0$, so that these partial derivatives can also be measured in parallel by post-processing measurements performed in a single basis. Continuing this pattern and defining the multi-index $\vec{\alpha}\in[n]^t$, one finds that the $t^{\text{th}}$ order partial derivatives are
\begin{align}
    \frac{\partial C^t}{\partial \theta_{\vec{\alpha}}} &= i^{t}2^t\bra{\psi_0} U^{\dagger}(\vec{\theta})\left(\prod_{\ell=1}^t G_{\alpha_\ell}\right)\mathcal{H}U(\vec{\theta})\ket{\psi_0}
\end{align}
if the $G_{\alpha_1},\cdots,G_{\alpha_t}$ all anticommute with $\mathcal{H}$, and zero otherwise. The non-zero $t^{\text{th}}$ order derivatives can therefore be evaluated by measuring the observables 
\begin{align}\label{fullops}
O_{\vec{\alpha}}=i^{t}2^t\left(\prod_{\ell=1}^t G_{\alpha_\ell}\right)\mathcal{H}.
\end{align} 
Once again, we find $[O_{\vec{\alpha}},O_{\vec{\alpha}'}]=0$ and so the $t^{\text{th}}$ order partial derivatives can also be obtained in parallel with variances scaling inversely with the number of shots to $C'$. Due to the factor $2^t$, the variance of the estimators will generally increase exponentially with the derivative order and thus in practice only reasonably low orders are within reach. However, this approach does enable approximate second-order optimisation with only a constant overhead in quantum resources compared to first-order methods.

Finally, we note that the form of \eqref{fullops} allows for an even stronger statement than Theorem \ref{thm:pds}.
\begin{corollary}
% Consider a parameterised commuting quantum circuit $C(\vec{\theta})$.
Consider a commuting-generator circuit $C(\vec{\theta})$ as in Theorem \ref{thm:grad}. There exist two circuits ${C}'_{\text{even}}$ and ${C}'_\text{odd}$ with the same number of qubits as $C$ such that all partial derivatives of even (resp.\ odd) order can be estimated by classically post-processing the circuit ${C}'_{\text{even}}$ (resp.\ ${C}'_{\text{odd}}$).
\end{corollary}
To see this, consider two observables $O_{\vec{\alpha}}\neq 0\neq O_{\vec{\alpha}'}$ with $\vec{\alpha}\in[n]^t$ and $\vec{\alpha}'\in[n]^{t'}$ where $t$ and $t'$ need not coincide.
Then we find $[O_{\vec{\alpha}},O_{\vec{\alpha}'}]=0$ if the parities of $t$ and $t'$ are the same ($t=t' \nospacemod 2$), and so all even (resp. odd) partial derivatives can be estimated in parallel.

\subsection{Fisher information matrix}
So far we have considered partial derivatives of cost functions as in Eq.~\eqref{ansatz}.
We now turn to the quantum Fisher information 
\begin{align}
    \mathcal{F}_{jk} &= \mathfrak{Re}\{
    \bra{\psi_0}[\partial_j U^\dagger(\vec\theta)] \partial_k U(\vec\theta) \ket{\psi_0}
    \} \\
    &- \bra{\psi_0}[\partial_jU^\dagger(\vec{\theta})] U(\vec{\theta})\ket{\psi_0}
    \bra{\psi_0}U^\dagger(\vec{\theta}) \partial_k U(\vec{\theta})\ket{\psi_0}\nonumber
\end{align}
of the parameterised quantum state;
it may be used for diagnoses of the prepared state itself \cite{Meyer2021fisherinformation,Toth_2014,fisher_info_universality}, or to compute the quantum natural gradient which can be used for training \cite{Stokes2020quantumnatural}.
For a commuting-generator circuit, we find that
\begin{align}
    [\partial_j U^\dagger(\vec{\theta)}]\partial_k U(\vec\theta) &= G_jG_k\\
    U^\dagger(\vec{\theta)} \partial_j U(\vec\theta) &= -i G_j.
\end{align}
This means that the Fisher information reduces to the covariance matrix of the generators in the encoding state $\ket{\psi_0}=V\ket{0}$:
\begin{align}
    \mathcal{F}_{jk} = \text{Cov}(G_j, G_k)_{\ket{\psi_0}}.
\end{align}
Note that all matrix entries can be measured in parallel because the generators commute. Furthermore, $\mathcal{F}$ does not depend on $\vec{\theta}$ but only on $V$, so that the same Fisher information matrix can be used throughout the optimisation.
Finally, we remark that some choices of encoding unitaries $V$ and gate generators even allow for a classical evaluation of $\mathcal{F}$, making the overhead of the quantum natural gradient purely classical. An example for this are the encoding and the generators used in Model A in Sec.\ \ref{sec:numerics}.
Overall, this makes estimating the natural gradient of commuting-generator circuits feasible in practice, provided that the number of parameters $n$ allows for inversion of an $n \times n$ matrix on a classical computer.

\begin{figure*}
    \centering
    \includegraphics[width=\textwidth]{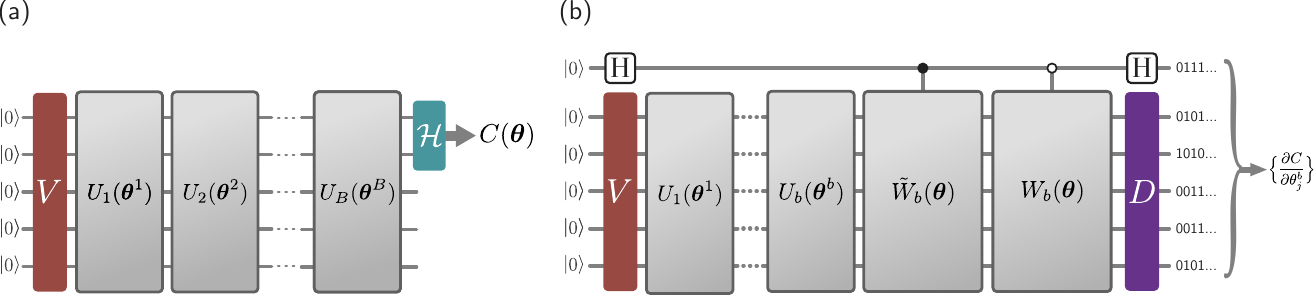}
    \caption{(a) The commuting-block circuit ansatz of Thm.\ \ref{thm:blocks}. The blocks $U_j(\vec{\theta}_j$) are commuting blocks with the same constraints as in Fig.\ \ref{fig:mainfig}. Generators between blocks have a fixed commutation relation: for any pair of blocks $U_j, U_k$ either (i) all generators from block $j$ commute with all generators from block $k$, or (ii) all generators from block $j$ anticommute with all generators from block $k$. (b) The circuit used to estimate the partial derivatives of those generators $G_j$ of block $b$ that anticommute with $\mathcal{H}$. Here $D$ is the unitary that diagonalises the operators $\{2iG_j\mathcal{H}\}$, and the unitaries $W_b$ and $\tilde{W}_b$ are defined in \eqref{Wbdef} and \eqref{Wbtildedef}. For generators $G_j$ that commute with $\mathcal{H}$, one replaces $W_b$ by $iW_b$ and $D$ diagonalises the operators $\{2G_j\mathcal{H}\}$. 
    }
    \label{fig:blocks}
\end{figure*}

\subsection{Simulability}\label{sec:simulate}
A natural question to ask is under what conditions commuting-generator circuits admit an efficient classical simulation. Since $V$ is arbitrary, one cannot expect to sample from the output distributions of such circuits in general. However, this holds even for $V=\mathbb{I}$. For example, by choosing a commuting circuit with Pauli generators in the $X$ basis, and an observable in the $Z$ basis, one arrives at a parameterisation of the class of IQP circuits \cite{bremner2016average,ni2012commuting}, for which sampling is known to be hard (up to non-collapse of the polynomial hierarchy) \cite{bremner2016average}. 

However, for the weaker task of expectation value estimation, commuting-Pauli-generator circuits can become classically tractable under certain conditions. More precisely, the results of \cite{ni2012commuting} imply the following. 

\begin{corollary}\label{thm:sim}
    Consider a commuting-Pauli-generator circuit with initial unitary $V$ and observable
    \begin{align}
        \mathcal{H}=F^\dagger Z \otimes \mathbb{I}\otimes\cdots\otimes\mathbb{I} F
    \end{align}
    for some Clifford unitary $F$. 
    Then if $FV\ket{0}=\ket{x}$ for some computational basis state $\ket{x}$, there is an efficient classical algorithm that estimates $C(\vec{\theta})$ to the same precision as the quantum circuit.
\end{corollary}
For a proof of the above, see App.\ \ref{app:simproof}. 

Note that the \textit{dynamical Lie algebra (DLA)} (see Sec.~\ref{sec:increasedexpressivity} for a definition) of any commuting circuit with polynomial depth is Abelian and has polynomial size.
Therefore, somewhat similar to the Corollary above, any input state and measurement observables with efficient classical representations that are compatible with a compressed representation of the DLA will lead to classical simulability as well~\cite{goh2023,somma2005quantum}.

\section{Commuting-block circuits}\label{sec:block}
We now present a larger class of circuits which retains the property of fast gradient estimation. These circuits have a similar structure to the commuting generator circuits, however the parameterised part of the circuit consists of blocks of commuting gates (see Fig.\ \ref{fig:blocks}). These blocks are such that while generators within a given block commute, generators between different blocks have a fixed commutation relation (either commutation or anticommutation). 
More precisely, if $\mathcal{G}_1 = \{G^1_j\}$ and $\mathcal{G}_2 = \{G^2_k\}$ are the generators from two distinct blocks $j,k$ then one has either 
\begin{align}
\left[G^1_j,G^2_k\right]=0\;\forall j,k \quad \text{  or  } \quad \left\{G^1_j,G^2_k\right\}=0\;\forall j,k. 
\end{align}
A general circuit in this class with $B$ blocks therefore has the form of \eqref{ansatz} with 
\begin{align}
U(\vec{\theta})= \prod_ j \exp(-iG^B_j\theta^B_j)\cdots \prod_ j \exp(-iG^1_j\theta^1_j)
\end{align}
with the blocks $\mathcal{G}_b = \{G_j^b\}$ respecting the aforementioned commutation structure. We will call such circuits 
% \emph{commuting-block quantum circuits}.
\emph{commuting-block circuits}.
Gradients of these circuits can be evaluated with a number of circuits that scales with the number of blocks rather than the number of parameters, in the following sense. 

\begin{theorem}\label{thm:blocks}
Consider an $N$-qubit commuting-block circuit with $B$ blocks. Then an unbiased estimate of the gradient $\nabla C(\vec{\theta})$ can be obtained by classically post-processing $2B-1$ circuits on $N+1$ qubits with increased depth. The variance of the estimator of each partial derivative scales as $\mathcal{O}(1/M)$ where $M$ is the number of shots used to evaluate each circuit.
\end{theorem}

We now explain the procedure to estimate the gradient; the full proof can be found in App.\ \ref{app:blockgrads}. To estimate the gradient of the $b^{\text{th}}$ block we use a technique based on linear combination of unitaries methods \cite{childs2012hamiltonian}. Since we focus on block $b$ we will drop the block index on the generators, so that $G_j\equiv G_j^b$. Define 
\begin{align}
\ket{\psi_b}=U_b(\vec{\theta}^b)\cdots U_2(\vec{\theta}^2)U_1(\vec{\theta}^1)V\ket{0},   
\end{align}
i.e.\ the state after the $b^{\text{th}}$ block, and 
\begin{align}\label{Wbdef}
    W_b(\vec{\theta}) = U_B(\vec{\theta}^B)\cdots U_{b+2}(\vec{\theta}^{b+2})U_{b+1}(\vec{\theta}^{b+1}).
\end{align}
We first define two sets of Hermitian observables: $\mathcal{O}_0 = \{O_j\} = \{2G_j\mathcal{H}\}$, for $j$ such that $G_j$ and $\mathcal{H}$ commute, and $\mathcal{O}_1 = \{O_j\} = \{2iG_j\mathcal{H}\}$, for $j$ such that $G_j$ and $\mathcal{H}$ anticommute.
In contrast to commuting-generator circuits, the generators in $\mathcal{O}_0$ contribute to the gradient although they commute with $\mathcal{H}$, which is due to the circuit $W_b$ applied after the $b^{\text{th}}$ block.
Note that all generators in either $\mathcal{O}_0$ or $\mathcal{O}_1$ mutually commute and so can be simultaneously diagonalised. We first focus on the partial derivatives corresponding to the indices $j$ appearing in $\mathcal{O}_1$. One implements the circuit shown in Fig.\ \ref{fig:blocks} (b) to produce the quantum state $\ket{\psi}$, where $D$ is the unitary that diagonalises the observables in $\mathcal{O}_1$, and $\tilde{W}_b$ is defined by 
\begin{align}\label{Wbtildedef}
    W_bG_j = G_j\tilde{W}_b.
\end{align}
For example, if generators in block $b$ anticommute with all generators in $W_b$ then $\tilde{W}_b$ has the same form as $W_b$ with $\vec{\theta}$ replaced by $-\vec{\theta}$. 
% In App.~\ref{app:blockgrads} we compute that measuring $Z\otimes O_j$ in the state $\ket{\phi_b}=H_\text{aux}\bar{c}W_bc\tilde{W}_b\ket{+}\ket{\psi_b}$--or its diagonalised equivalent in the state $D\ket{\phi_b}$--yields the required derivative:
% One finds that the state without the final unitary $D$ is such that
For this circuit one finds 
\begin{align}
    \bra{\psi} Z\otimes DO_jD^\dagger\ket{\psi}=\partial_j C(\vec{\theta}).
\end{align}
Since $D$ diagonalises the $O_j$, the partial derivatives are obtained by sampling the circuit in the computational basis via post-processing analogous to \eqref{postprocess}. For $\mathcal{O}_0$ the process is the same, however one replaces $W_b$ by $iW_b$ in the circuit and uses a unitary $D$ that diagonalises the observables in $\mathcal{O}_0$. For each block we therefore require two circuits, however the generators that commute with $\mathcal{H}$ in the final block have zero gradient; hence $2B-1$ circuits suffice. 

% To estimate the gradient of the $b^{\text{th}}$ block we use a technique based on a linear combination of unitaries circuit (see App.\ \ref{app:blockgrads} for a full description). For clarity, here we focus on the specific case where all generators in block $b$ anticommute with the observable $\mathcal{H}$.  Define \begin{align}
% \ket{\psi_b}=U_b(\vec{\theta})\cdots U_2(\vec{\theta})U_1(\vec{\theta})V\ket{0}    
% \end{align}
% and 
% \begin{align}
%     W_b = U_B(\vec{\theta})\cdots U_{b+2}(\vec{\theta})U_{b+1}(\vec{\theta}).
% \end{align}
% To get the gradient of this block one implements the three circuits shown in Fig.\ \ref{fig:blocks}, and combines their outcomes. Note that although the linear combination of unitaries method is often used to probabilistically implement a transformation, the method we present to estimate the gradient is deterministic; i.e. all outcomes contribute to the value of the gradient.

\begin{figure*}
    \centering
    \includegraphics[scale=0.8]{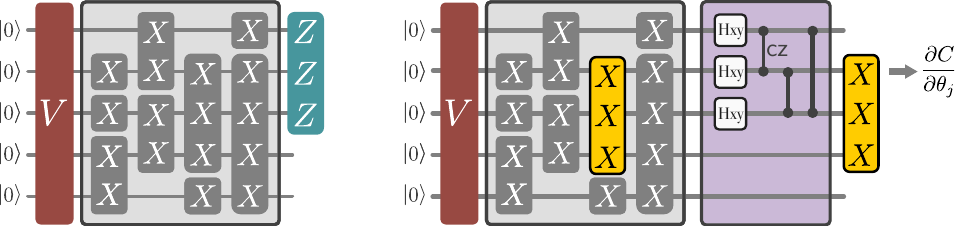}
    \caption{An instance of an $X$-generator circuit for 5 qubits (left) and the corresponding circuit used for gradient estimation (right). The parameterised gates are generated by products of Pauli $X$ operators on different subsets of qubits and  the observable is a product of Pauli $Z$ operators on a subset of qubits. To estimate the gradient, an additional circuit $D$ is added, which involves performing a controlled $Z$ gate between all qubits on which the observable acts non-trivially. Here $H_{xy}$ is the single-qubit Hadamard unitary that switches between the $X$ and $Y$ bases. To estimate the gradient of one of the gates (highlighted in yellow), one evaluates the expectation value of the generator at the output of the circuit. Since all generators are products of $X$, one can estimate the gradient in parallel by measuring each qubit in the $X$ basis.}
    \label{fig:xcircuit}
\end{figure*}

\subsection{Increased expressivity with commuting-block circuits}\label{sec:increasedexpressivity}
A natural question to ask is whether commuting-block circuits have increased expressivity compared to the single block circuits of Fig.\ \ref{fig:mainfig}. This is indeed the case. To quantify expressivity we will use the \emph{dynamical Lie algebra} \cite{ragoneunified} of the gate generators, defined as 
\begin{align}\label{DLA}
    \text{DLA}[\{G_j\}] = \text{span}_\mathbb{R}\langle\{iG_j\}\rangle_{\text{Lie}}
\end{align}
where $\text{span}_\mathbb{R} S$ is the real vector space spanned by the elements of $S$ and $\langle \mathcal{S} \rangle_{\text{Lie}}$ is the Lie closure: the set of nested commutators of operators in $\mathcal{S}$. As a simple example, a single-qubit ansatz with the Pauli $X$ and $Y$ operators as generators has the DLA $=\text{span}_\mathbb{R}\{iX,iY,iZ\}$ since $[iX,iY]=-2iZ$. The class of unitaries that can be realised by sequential uses of gates with generators $\{G_j\}$ is given by \cite{controlbook,larocca2022diagnosing}
\begin{align}\label{UDLA}
    \mathcal{U} = \{e^{X} \vert X\in \text{DLA}[\{G_j\}] \}.
\end{align}
Thus, the larger the dimension of the DLA, the larger the expressivity of the ansatz. Note that the dimension of the DLA cannot be larger than $4^N$, since this is the dimension of the space of Hermitian operators acting on $N$ qubits. 
%
% \begin{align}
%     \mathcal{U} = \{e^{X} \vert X\in \text{DLA}[\{G_j\}] \}.
% \end{align}
%
% Thus, the larger the DLA, the larger the expressivity of the ansatz. 

For a single block, since all generators commute, the dimension of the DLA is given by the dimension of the span of the generators, and therefore cannot be greater than $2^N$. This dimension can be achieved for example by considering the $2^N$ generators comprised of tensor products of $Z$ and $\mathbb{I}$. 
To show that we can go beyond this limit, consider a commuting-block ansatz consisting of sequential uses of the following two blocks. The first block is comprised of $2^N/2$ gates that are generated by $G_j\in\{\mathbb{I}, Z\}^N$, where we consider operators with an odd numbers of $Z$s only. The second block contains a single gate generated by $X^{\otimes N}$.
These blocks satisfy the condition of mutual anticommutativity. A small calculation shows that the DLA is given by linear combinations of (i) the generators of the first block (of which there are $2^N/2$) and (ii) all full-weight Pauli tensors of $X$ and $Y$ operators (of which there are $2^N$). Thus we have a DLA with dimension $\frac{3}{2}\cdot 2^N$, which is larger than the maximal DLA for a single block.

\subsection{Gradient scaling}
The ability to train parameterised quantum circuits under random initialisation of the parameters is closely connected to the phenomenon of barren plateaus \cite{mcclean2018barren}, whereby the magnitudes of cost function partial derivatives decrease exponentially with the number of qubits. Recent work has proven that if either the initial density matrix or cost observable are in the DLA of the gate generators, then the scaling of the partial derivatives is closely tied to the inverse of the dimension of the algebra \cite{ragoneunified, fontanaadjoint}. 

For commuting-generator circuits, the DLA is simply the span of $\{G_j\}\cup \mathbb{I}$, which thus has a polynomially bounded dimension for polynomially sized circuits. Unfortunately, the results of \cite{ragoneunified, fontanaadjoint} cannot be applied to commuting generator circuits however, since the requirement that the input state or observable be in the DLA implies commutation with all generators, and the gradients of such circuits will therefore vanish. The interesting cases for these circuits are therefore when both the input state and observable lie outside of the DLA, however this is not covered by the current theory. 

For commuting-block circuits the DLA can contain non-commuting basis elements, and so non-trivial circuit structures that are free from barren plateaus can in principle be constructed, provided that the dimension of the DLA is kept polynomial (here, results from \cite{wiersema2023classification} may be useful). Nevertheless, one must be careful, since using an observable in the DLA can often render the circuit simulable \cite{goh2023}. One may therefore need to find a `sweet spot' that avoids the criteria in \cite{goh2023} for efficient simulation, but retains non-vanishing gradients from the results of \cite{ragoneunified, fontanaadjoint}. 

Finally, we stress that although barren plateaus are often viewed as a practical barrier to training, their existence does not necessarily prohibit efficient training via strategies that do not initialise parameters uniformly at random. For example, alternative initialisation strategies \cite{grant2019, sack2022, zhang2022, volkoff2021}, layer-wise training \cite{skolik2021} or classical pre-training \cite{rudolph2023, simon2023} may be effective approaches to optimizing such circuits in practice.

\section{Explicit constructions of commuting circuits}\label{sec:examples}
In this section we present two constructions of commuting generator circuits, and detail the explicit unitaries required for gradient estimation. As we will see, both these constructions allow for backpropagation scaling for gradient estimation. 
    
\subsection{\texorpdfstring{$X$}{X}-generator ansatz}\label{sec:xgen}
For this ansatz the generators are diagonal in the $X$ basis:
\begin{align}\label{GjX}
    G_{j} = G_{j, 1}\otimes\cdots\otimes G_{j, N}  \in \{\mathbb{I},X\}^{\otimes N}
\end{align}
and the possible observables are diagonal in the $Z$ basis:
\begin{align}
    \mathcal{H} \in \{\mathbb{I},Z\}^{\otimes N}. 
\end{align}
Recall that to estimate the gradient we need to evaluate observables $O_j=2iG_j\mathcal{H}$ for all $G_j$ that anticommute with $\mathcal{H}$. We begin with the simple case where the observable is
\begin{align}
    \mathcal{H} = Z\otimes\mathbb{I}\otimes\cdots\otimes\mathbb{I}. 
\end{align}
We need only consider generators that feature an $X$ on the first qubit. The corresponding observables $O_j$ take the form 
\begin{align}\label{ojsX}
    O_j = 2 Y\otimes G_{j, 2}\otimes \cdots \otimes G_{j, N}
\end{align}
and mutually commute as expected. In this case, simultaneous diagonalisation is easy since the $O_j$ are diagonal in the $Y\otimes X\otimes \cdots\otimes X$ product basis. To obtain estimates for all partial derivatives one samples the $N$ qubits in this basis and then evaluates expectation values of the relevant subsets of qubits given by \eqref{ojsX}.

When there is more than one $Z$ operator appearing in $\mathcal{H}$ the situation is a little more complicated. Let us assume that $\mathcal{H}$ is of the form
\begin{align}\label{zobs}
    Z\otimes \cdots \otimes Z \otimes\mathbb{I}\otimes \cdots \otimes\mathbb{I},
\end{align}
i.e. the $Z$ operators appear on the first $m$ qubits only. Due to the symmetry of the ansatz, other cases can be handled by permuting the qubits in the corresponding circuits. 

\begin{figure*}
    \centering
    \includegraphics[scale=0.8]{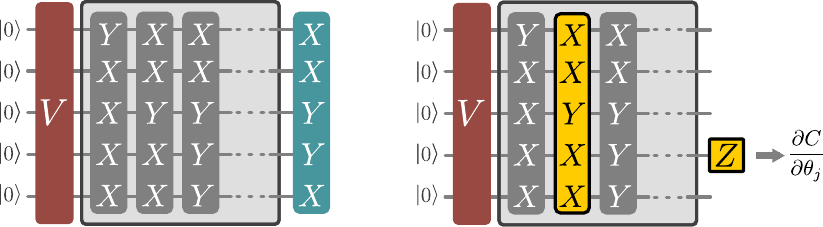}
    \caption{A commuting-generator circuit family with nonlocal generators and Hamiltonian (left).
    The similarity in structure between the Hamiltonian and the generators allows to measure all gradient entries in parallel without an additional basis change, simply by measuring in the computational basis and post-processing the samples into expectation values of products of Pauli $Z$ operators (right).}
    \label{fig:nlcircuit}
\end{figure*}

% Let us denote the number of $X$ operators of $G_j$ on the first $m$ qubits as $m_j$. Since we need only consider generators that anticommute with $\mathcal{H}$, we have that $m_j$ is odd and the $O_j$ take the form 
% %
% \begin{align}\label{Oj1}
%     O_j = 2s_j P_{j,1}\otimes \cdots \otimes  P_{j, m} \otimes G_{j, m+1}\otimes \cdots \otimes G_{j, N},
% \end{align}
% %
% where $P_{j, r}\in{Z,Y}$ with an odd number of $Y$s. The sign factor takes the form
% %
% \begin{align}\label{factor}
%         s_j
%         = i^{m_j-1}
%         &=\begin{cases}
%             1 & \text{ for } m_j \nospacemod 4 = 1\\
%             -1 &  \text{ for } m_j \nospacemod 4 = 3
%         \end{cases}.
% \end{align}
We now need to diagonalise the operators $O_j=2iG_j\mathcal{H}$ (for $\{G_j,\mathcal{H}\}=0$), however the diagonalising unitary is no longer a tensor product of single-qubit unitaries. In App.\ \ref{app:xgrads} we show that the circuit of Fig.\ \ref{fig:xcircuit} diagonalises the $O_j$ in the product $X$ basis. This circuit involves applying an $x-y$ Hadamard on the first $m$ qubits followed by a controlled-Z between every pair of the first $m$ qubits. More precisely, the effect of the circuit is to map the $O_j$ back to the $G_j$:
\begin{align}
    D O_j D^\dagger = 2G_j.
\end{align}
Thus, the partial derivative $\partial_j C(\vec{\theta})$ can be obtained by estimating the expectation value of $G_j$ at the output of the circuit of Fig.\ \ref{fig:xcircuit}. Since the $\{G_j\}$ take the form \eqref{GjX}, one simply needs to measure each qubit in the $X$ basis and construct the expectation value from the relevant subset.

% These operators are diagonalised by the circuit $D$ shown in Fig.\ \ref{fig:Xdiag}, which requires applying a controlled-Z gate between all pairs of the first $m$ qubits. %
% %Denote by $S_j\subset\{1,2,\cdots,n\}$ the subset of qubits in which an $X$ appears in $G_j$ and $\vert S_j \vert$ the number of elements in $S_j$. 
% One finds
% %
% \begin{align}
%     D O_j D^\dagger = 2(\bigotimes_i^n H)G_j(\bigotimes_i^n H),outperforms
% \end{align}
% %
% i.e. the digaonalised operator has the same structure as $G_j$ with $X$ replaced by $Z$. Denoting by $S_j\subset\{1,2,\cdots,n\}$ the subset of qubits in which an $X$ appears in $G_j$, partial derivatives can thus be evaluated as 
% %
% \begin{align}
%     % \frac{\partial C}{\partial \theta_j}=(-1)^{(\frac{\vert S_j \vert-1}{2})^2}\mathbb{E}\Big[\prod_{i\in S_j} z_{i}\Big]
%     \frac{\partial C}{\partial \theta_j}=\mathbb{E}\Big[\prod_{i\in S_j} z_{i}\Big]
% \end{align}
% where $z_i$ is the $i^{\text{th}}$ output bit of the circuit. That is, the partial derivative of the generator $G_j$ corresponds to an expectation value of the subset of qubits on which it acts non-trivially; 

The second-order derivatives are given by
\begin{align}
    \partial_{ij}C=\bra{0}V^\dagger U^{\dagger}(\vec{\theta})O_{jk}U(\vec{\theta})V\ket{0}
\end{align}
with $O_{jk} = -4G_jG_k\mathcal{H}$ such that both $G_j$ and $G_k$ anticommute with $\mathcal{H}$. The $\{O_{jk}\}$ can be diagonalised by a similar circuit to $D$: one applies a $z-y$ Hadamard on the first qubit, followed by $D$ (see App.\ \ref{app:xgrads}). With this one finds
\begin{align}\label{xgjgk}
    D H_{zy}^{(1)} O_{jk} H_{zy}^{(1)} D^\dagger = (-1)^{\beta_1} 4 X^{(1)}G_jG_k,
\end{align}
where $\beta_1=0$ if $G_{j,1}G_{k,1}=X$ and $1$ otherwise. The second-order partial derivatives can therefore be obtained by replacing $D$ by $D H_{zy}^{(1)}$ in Fig.\ \ref{fig:xcircuit}, measuring in the $X$ basis and estimating the expectation values of the relevant subset of qubits given by \eqref{xgjgk}. 

Note that, if we restrict to local observables (i.e.\ $m$ is upper bounded by a constant for all $N$), then the diagonalising unitaries for both the gradient and second order partial derivatives are of constant depth, and we achieve backpropagation scaling for the gradient estimation. 

\subsection{Circuits with nonlocal generators}
Here we give another construction that uses fully nonlocal generators. While it generally comes at the cost of larger circuit depths, we will see that this structure allows the gradient of a sum of observables to be estimated from a single circuit. 
The generators we consider take the form 
\begin{align}
    G_j \in \{X,Y\}^{\otimes N}
\end{align}
with an odd number of $Y$'s in each generator. The observables take the same form
\begin{align}\label{nlobs}
    \mathcal{H} \in \{X,Y\}^{\otimes N}
\end{align}
but with an even number of $Y$'s. One sees that the observables $O_j=2iG_j\mathcal{H}$ are up to a sign factor given by products of $Z$ operators on those qubits on which $G_j$ and $\mathcal{H}$ differ. The gradient of $\braket{\mathcal{H}}$ can therefore be estimated by sampling each qubit in the $Z$ basis and estimating the relevant expectation value. This holds true for any observable $\mathcal{H}_p$ of the form \eqref{nlobs}. Thus we can in fact estimate the gradient of any operator
\begin{align}
    \mathcal{H} = \sum_p h_p\mathcal{H}_p
\end{align}
by measuring the same circuit in the $Z$ basis and postprocessing. Since the depth of the gradient evaluation is the same as the model, we achieve backpropagation scaling. 

% by post-processing a single circuit. We remark that since the observables $\mathcal{H}_p$ commute, the parameter-shit rule can also benefit from this sort of parallelisation over sums of observables by measuring in a basis in which the $\mathcal{H}_p$ are diagonal. This requires performing an additional diagonalising circuit however, which is not the case for the shown circuit. 

\begin{figure*}
    \centering
\includegraphics[width=\textwidth]{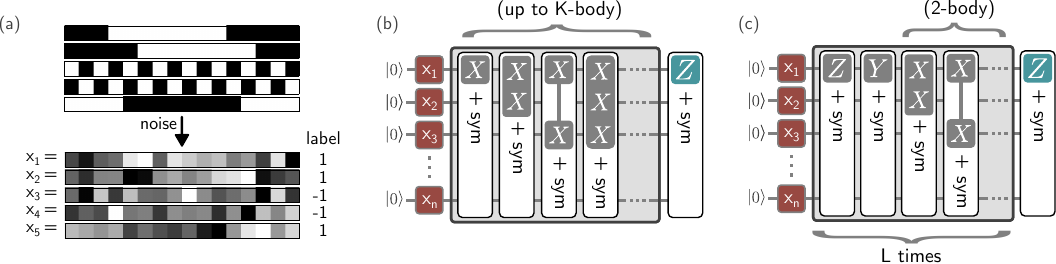}
    \caption{(a) We generate two types of 16-dimensional vectors corresponding to either bars (label 1) or dots (label 2). Independent Gaussian noise is added to generate the input data for the classification task. Note that the labels of the data set are invariant to translations of the elements of the vectors. 
    (b--c) The two translation equivariant models we benchmark for this problem. Here `+ sym' denotes symmetrisation of the generator over 1D translations, e.g.\ $X_1$ + sym $=X_1+X_2\cdots +X_n$.
    The model in (b) features $X$ generators only, and can therefore exploit the parallel gradient evaluation described in Sec.\ \ref{sec:xgen}. The model in (c) is a translation equivariant version of the model in \cite{schatzki2022theoretical} and features non-commuting generators.
    }
    \label{fig:models}
\end{figure*}

\section{Numerical study: learning translationally invariant data}\label{sec:numerics}
Although the circuits introduced in this work require much fewer circuit shots to optimise, it remains unclear if their limited structure makes them suitable in practice. In this section we investigate this by using a machine learning problem based on classifying 1D translationally invariant data. We first construct a 1D translation equivariant quantum model based on the $X$-generator ansatz of Sec.\ \ref{sec:xgen} that is tailored for learning data with this structure. We then train this model (Model~A) and compare it against three other models: a 1D translation equivariant quantum model with non-commuting generators (Model~B), inspired by a recently introduced permutation equivariant model \cite{schatzki2022theoretical}; a quantum convolutional neural network \cite{qconv,hur2022quantum} (Model~C); and a simple separable quantum model (Model~D). As we will see, the commuting circuit model (Model A) performs the best of the four models whilst requiring much fewer circuit shots to train. This suggests that commuting-generator circuits may generally perform as well or even better than other proposed models from the literature, and could therefore represent a powerful model class for machine learning tasks. Code to reproduce our numerical results is available online \cite{backpropscalingrepo}.

\subsection{The learning problem: learning bars and dots}
The dataset we consider can be seen as a simple 1D version of the bars and stripes data set that we call `bars and dots' (see Fig.\ \ref{fig:models} (a)). Each data point $\vec{x}_i$ is a $d$-dimensional real vector, which corresponds to a bar (class 1) or dots (class 2). A bar vector is constructed by setting the values of $\lfloor d/2 \rfloor$ neighbouring elements of the vector to $+1$ and the remaining values to $-1$. A dots vector corresponds to alternating $\pm1$ values. The dataset is created by setting $d=16$, sampling random bars and dots vectors, and then adding independent Gaussian noise (with variance 1) to each element of the vector. The learning problem is a simple binary classification problem to predict whether a given vector belongs to the bar or dots class (corresponding to a label $y_i=\pm1$), and the figure of merit we consider is the usual class prediction accuracy on a test dataset consisting of 100 input vectors. During the training phase, we generate a set of 1000 input vectors and train each model with the adam optimiser using batches of 20 input vectors per update, an initial learning rate of $0.01$ and gate parameters initialised uniformly random in $[0,2\pi]$. All models are trained using the binary cross entropy loss
\begin{align}
    - \sum_i \log P(y_i\vert \vec{x}_i, \vec{\theta})
\end{align}
where the probability $P(+1\vert \vec{x}_i, \vec{\theta})$ is given by $\sigma(6\langle \mathcal{H} \rangle)$, where $\sigma$ is the logistic function and $\mathcal{H}$ is normalised such that $\vert\langle \mathcal{H} \rangle\vert\leq 1 $. In practice, we observed that this gives better results than taking for example $P(+1\vert \vec{x}_i, \vec{\theta})=(1+\mathcal{H})/2$. 

\subsection{Quantum models}
Each model corresponds to a quantum circuit with $N=16$ qubits, where the data is encoded into the circuit via single-qubit Pauli $Y$ rotations given by the data encoding unitary
\begin{align}
    V_{\vec{x}} = \bigotimes_{r=1}^d 
    % e^{-\frac{x^{(j)}}{4}Y}
    % e^{-\frac{i}{4}x^{(r)} Y_r}
    \exp\left(-\frac{i}{4}x^{(r)} Y_r\right)
\end{align}
with $x^{(r)}$ the $r^{\text{th}}$ entry of $\vec{x}$. Note that we have chosen a specific scaling of the data in the above; in practice we found that this generally gives good results compared to other scalings, however we have not performed a detailed investigation into the effect of different data scalings on each model. After this data encoding layer, each model consists of a parameterised unitary followed by measurement of an observable $\mathcal{H}$, where we use $\text{sign}(\langle\mathcal{H}\rangle)$ as our class prediction label. The parameterised unitaries for each model are as follows. 

\emph{Model A: Equivariant $X$-generator circuit}---This model is an instance of the $X$-generator ansatz of Sec.\ \ref{sec:xgen}, and is depicted in Fig.\ \ref{fig:models}(b). To construct a model with equivariant layers, we need to consider gate generators that commute with the symmetry group that corresponds to the label invariance of the data \cite{nguyen2022, ragone2022representation}. 
From our choice of data encoding, the translation symmetry of the data is represented by translations of the qubit subsystems; i.e.\ translating the elements of the input vector $\vec{x}$ results in a new data encoding unitary $SV_{\vec{x}}$, where $S$ is an operator that cyclicly permutes the qubits. We therefore consider generators that are symmetric with respect to this symmetry group. The simplest of these corresponds to the operator
\begin{align}
    X_1 + X_2 + \cdots X_d = \text{sym}(X_1),
\end{align}
where $\text{sym}(X_1)$ is the twirling operation \cite{meyer2023exploiting,nguyen2022} that symmetrises a given operator with respect to 1D cylic permutations. Two-body generators are therefore given by $\text{sym}(X_1X_{r})$ for different choices of $r$. The ansatz uses all $k$-body generators up to some fixed $K$, which we leave unspecified for now. By using an observable that also respects this symmetry, we arrive at an equivariant model whose label prediction is invariant to the data symmetry. This follows from
\begin{multline}
    \bra{0}V_{\vec{x}}^\dagger S^\dagger U(\theta)^\dagger \mathcal{H} U(\theta) S V_{\vec{x}}\ket{0}  \\ = \bra{0}V_{\vec{x}}^\dagger U(\theta)^\dagger \mathcal{H} U(\theta) V_{\vec{x}}\ket{0},
\end{multline}
where we have used the fact that $S$ commutes with $\mathcal{H}$ and $U(\theta)$ by construction. To this end, we consider the observable 

\begin{align}
    \mathcal{H} = Z_1 + Z_2 + \cdots + Z_d.
\end{align}
Note that the gradient for each $Z_r$ can be estimated in parallel via the method of Sec.\ \ref{sec:xgen}, as will be described shortly in Sec.\ \ref{sec:modelshots}.

\begin{figure*}
    \centering
    \includegraphics[width=\textwidth]{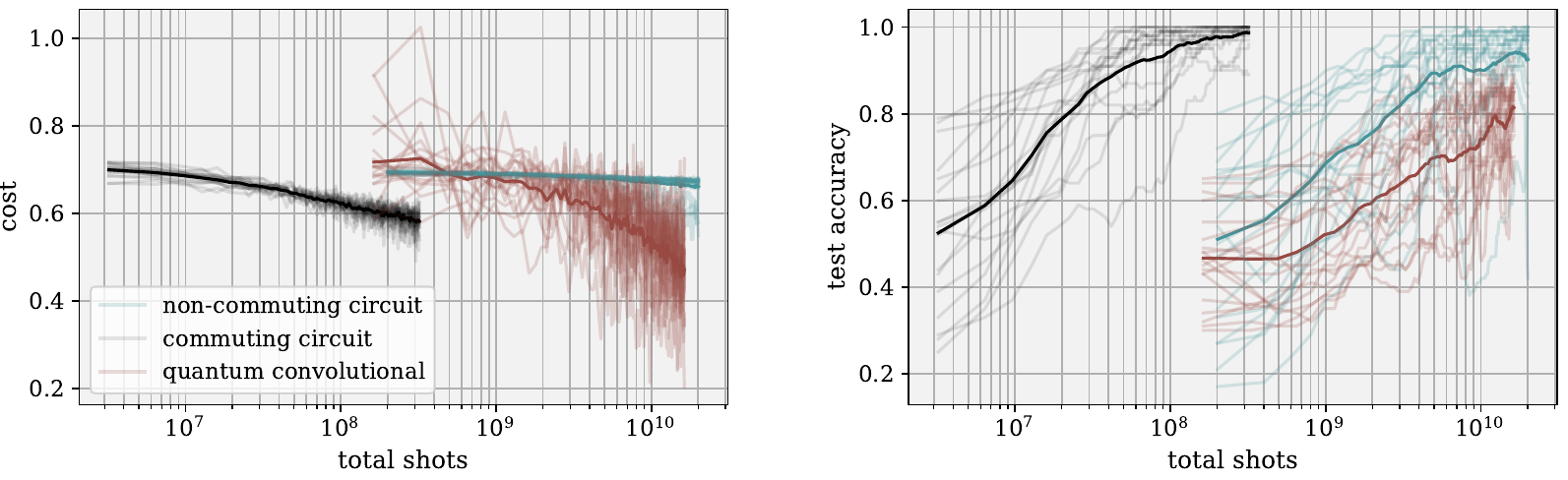}
    \caption{Numerical results for the `bars and dots' learning problem using 16-qubit models and numerically exact gradient evaluation (simulated using PennyLane \cite{bergholm2018pennylane} and JAX \cite{jax2018github}). All models are trained using the cross entropy loss and we assume $M=10000$ shots per circuit used for gradient estimation. The plots show the results after $100$ gradient steps using the adam optimiser, for 19 trials per model (average values given by thick lines). The highest average accuracy is achieved by the commuting-generator model, despite requiring two orders of magnitude fewer shots.}\label{fig:numerics1}
\end{figure*}

\emph{Model B: Equivariant model with non-commuting generators}---This model (see Fig.\ \ref{fig:models}(c)) can be seen as a translation equivariant analogue of the permutation equivariant model presented in \cite{schatzki2022theoretical}. The model consists of $L$ layers where the generators in each layer are given by $\text{sym}(Z_1)$, $\text{sym}(Y_1)$ and $\text{sym}(X_1X_{r})$. The final observable $\mathcal{H}=\text{sym}(Z_1)$ is the same as for Model~A. Since this model is not of the form of a commuting-generator or commuting-block circuit, gradient evaluation will be assumed to be via the parameter-shift rule.  

\emph{Model C: Quantum convolutional neural network}--We also train a quantum convolutional quantum neural network \cite{qconv,hur2022quantum}. We construct the model based on the architecture in \cite{hur2022quantum}, using a 10-parameter convolutional circuit (circuit 7 of Fig.\ 2 therein). Here the observable is given by $\mathcal{H}=Z_d$. As with Model~B, gradients are assumed to be evaluated via the parameter-shift rule. 

\emph{Model D: Separable quantum model}---As a sanity check, we also train a fully separable quantum model. This model has $3$ trainable parameters per qubit that are used to optimise individual single-qubit rotations via the parameter-shift method. The observable is $\mathcal{H}=\text{sym}(Z_1)$.

\subsection{Calculating the required shots}\label{sec:modelshots}
Since we are interested in the number of shots used to train the models, we would ideally estimate gradients using a finite number of shots from each quantum circuit. 
% However, since just-in-time compilation is currently not supported for gradient evaluation of finite-shot devices in PennyLane, in practice this approach becomes very resource intensive at larger qubit numbers. 
However, the classical simulation of finite-shot gradients becomes very resource intensive at larger qubit numbers, making this approach inconveniently expensive.
For this reason we use numerically exact statevector simulations and automatic differentiation, which returns the exact gradient and can in principle result in different behaviour compared to a finite-shot scenario. To account for this, we add independent Gaussian noise with zero mean and standard deviation $\epsilon= 0.1$ to the gradient vector before updating the parameters. This roughly approximates the shot noise that one would encounter from sampling $M=1/\epsilon^2=10000$ shots from each quantum circuit involved in gradient evaluation.

% To account for this, we work under the assumption that the number of shots $M$ taken from each circuit is large enough to ensure that the estimated gradient matches the exact gradient to sufficient precision. In this way we use the exact batch gradient for updates, and add $M$ shots for each circuit that would be used in the corresponding gradient method. 

We now proceed to calculate the number of shots needed to estimate the gradient for a fixed input $\vec{x}$ for each of the four models. For additional details of the calculation, also see App.\ \ref{app:shot_estimation}. For Model A, the gradient for each $Z_r$ can be estimated using the parallel gradient method of Sec.~\ref{sec:xgen}. For example, for $r=1$, if we consider the generator $\text{sym}(X_1)$ with parameter $\theta_1$ then 
\begin{align}
    \frac{\partial C}{\partial \theta_1} &= \bra{0}V_{\vec{x}}^\dagger U^\dagger(\vec{\theta})[i\,\text{sym}(X_1),Z_1] U(\vec{\theta})V_{\vec{x}}\ket{0}\nonumber\\
    &=\bra{0}V_{\vec{x}}^\dagger U^\dagger(\vec{\theta})\left(O_1+\cdots + O_d\right) U(\vec{\theta})V_{\vec{x}}\ket{0},
\end{align}
where $O_j=i[X_j,Z_1]$. It follows that each component of the gradient is obtained by summing contributions from operators $O_j$ of the form \eqref{ojsX}, which we have seen can be performed in parallel. Since we need to do this for each $Z_r$ in $\mathcal{H}$, the gradient evaluation requires $M\cdot d$ shots in total. 

For Model B, similarly to above, considering the first generator $\text{sym}(Z_1)$ with parameter $\theta_1$ we find
\begin{align}
    \frac{\partial C}{\partial \theta_1} =& i\bra{0}V_{\vec{x}}^\dagger \text{sym}(Z_1)U^\dagger(\vec{\theta})\mathcal{H}U(\vec{\theta})V_{\vec{x}}\ket{0} \nonumber\\ &\quad- i\bra{0}V_{\vec{x}}^\dagger U^\dagger(\vec{\theta})\mathcal{H}U(\vec{\theta})\text{sym}(Z_1)V_{\vec{x}}\ket{0} 
    % \nonumber\\= 
    % &\sum_r i\bra{0}V_{\vec{x}}^\dagger Z_rU^\dagger(\vec{\theta})\mathcal{H}U(\vec{\theta})V_{\vec{x}}\ket{0} \nonumber\\ &\quad- i\bra{0}V_{\vec{x}}^\dagger U^\dagger(\vec{\theta})\mathcal{H}U(\vec{\theta})Z_r V_{\vec{x}}\ket{0}
\end{align}
which is equivalent to a sum of $d$ partial derivatives for gates with generators $Z_r$ by expanding the sum in $\text{sym}(Z_1)$. Via the parameter-shift rule we therefore need to evaluate two circuits per Pauli generator used in the circuit.
Note that since one measures $\mathcal{H}$ directly here, we do not need to evaluate a different circuit for each $Z_r$ appearing in $\mathcal{H}$, unlike in the parallel method. The total number of shots required is $M_B=M\left[d^2(L-1)+3d(L+1)-2\right]$ for $L$ layers on $d$ qubits (see App.\ \ref{app:shot_estimation} for details).

For Model C, we need to be more careful, because not all gates are generated by Pauli words, since the convolutional network also uses controlled Pauli rotations. We compute the required shots to be $M_C = M(56d-80)$.

Finally, for Model D the gradient of each $Z_r$ depends only on qubit $r$ because the circuit is separable. Thus we can use the parameter-shift rule in parallel and a full gradient evaluation requires only $6M$ shots.

We see that the parallel method scales with $d$, whereas the parameter-shift rule scales with the number of Pauli generators: in a regime in which the number of parameters or the number of gates is much larger than the number of qubits, this can have a dramatic effect on shot efficiency. In our numerical experiment, Model B systematically suffers from this because it uses $\mathcal{O}(Ld^2)$ gates, whereas Model C merely suffers from a constant-factor overhead of $56$, which is significantly larger than $d=16$.
This means that Model C asymptotically requires the same number of shots as Model A, namely $\mathcal{O}(Md)$, however at the price of being limited to $\mathcal{O}(d)$ parameters. In Table \ref{tab:model_specs} we show the number of parameters and circuits for each model for the case $d=16$ considered here. 

Finally, we note that in the case of Model A, since gradients are estimated in parallel, correlations could exist between measurement outcomes that correlate the noise in the gradient vector in undesirable ways, although this will be less of a problem the higher the precision of the estimate. To investigate this, in App.\ \ref{app:finiteshot} we present results for a 6-qubit version of Model A, using a complete finite shot analysis with $500$ shots per circuit, and observe that the behaviour of the model closely matches that of the exact gradient.

\section{Results}

In Fig.\ \ref{fig:numerics1} (a) we show the results for the 16-dimensional bars and dots problem, considering 19 trials for each model, and 100 training steps for each trial.
The results for Model D can be found in Fig.\ \ref{fig:modelD} in the appendix; we do not include them here since the model performs very poorly compared to the others. 
For Model A, we consider a model containing all generators up to $3$-body interactions (so $K=3$), and for Model B we set the number of layers to $L=4$. This ensures that the models have approximately the same number of trainable parameters. In Tab.\ \ref{tab:model_specs} we show the number of parameters and the number of circuits required for gradient evaluation.

\begin{table}
\begin{center}
\resizebox{\columnwidth}{!}{%
\begin{tabular}{ |c|c|c| } 
\hline
Model & \# parameters $n$ & \# circuits \\
\hline
A (commuting) & 44 & 16 \\ 
B (non-commuting) & 40 & 1006 \\
C (q. convolutional) & 48 & 816 \\
D (separable) & 48 & 6 \\
\hline
\end{tabular}
}
\end{center}
\caption{Specifications and resource requirements for each model for the 16-qubit classification problem we consider. $n$ is the number of trainable parameters, and \# circuits is the number of unique circuits that need to be evaluated in order to estimate the gradient for a given input $\vec{x}$.}  \label{tab:model_specs}
\end{table}

The horizontal axes of Fig.\ \ref{fig:numerics1} show the (cumulative) number of shots used for the training, computed as described above, and assuming $M=10000$ shots per circuit. As expected, Model A requires dramatically fewer shots, and completes the 100 gradient steps before the first gradient update of the other models is available. For this problem, this translates to a roughly two orders of magnitude reduction in circuit shots. 

The training dynamics of the three models vary significantly, with Model C exhibiting a strongly fluctuating batch loss. Interestingly, this model achieves a lower average batch loss than the other models, however this is not in contradiction to the poorer test accuracy. Consider for example a model whose label prediction probabilities are close to $\frac{1}{2}$ on all inputs, but such that $\text{sign} \langle \mathcal{H} \rangle$ always returns the correct label. This model will have perfect accuracy, however the expected cross entropy on any dataset can be arbitrarily close to $-\log(\frac{1}{2})\approx 0.69$. Essentially, a similar effect is at play here; Model C is more confident in its predictions than Model A and B (resulting in a lower cross entropy), however its prediction $\text{sign} \langle \mathcal{H} \rangle$ is more likely to be incorrect.

Both equivariant models (A and B) exhibit smoother training dynamics and achieve much higher test accuracies. This suggests that equivariant architectures of the form presented here may be better suited to learn translation invariant data than previous proposals of quantum convolutional models (however we stress that a more involved hyperparameter optimisation and longer training times would be required to make a conclusive statement). Most importantly, the restriction to commuting generators appears not to compromise the performance of Model A, which even achieves the highest average test accuracy of all models. We expect that the flatter training curve of Model B is a reflection of a flatter cost landscape for this model; this may be linked to the fact that model has a larger DLA than Model A. 

\section{Discussion}
Although the results of Sec.\ \ref{sec:numerics} provide some indication that commuting-generator circuits can perform well at tasks of interest, it remains to be seen whether their reduced expressivity compared to general ans\"{a}tze remains sufficient to expect advantages over classical models when tested at scale. An important route to clarify this issue would be to understand the limits of expressivity of commuting-block circuits. As we have seen in Sec.\ \ref{sec:increasedexpressivity}, increasing the number of blocks can result in a larger expressivity compared to single-block ans\"{a}tze, however it remains unclear how far one can push this under the constraints on the circuit generators. Since these constraints involve mutual anticommutation between blocks, a natural starting point could be to consider maximal sets of mutually anticommuting Pauli tensors, of which it is known that there cannot be more that $2N+1$ \cite{sarkar2021sets}. Such a set can be found for example by considering the Jordan-Wigner form of a set of Majorana operators, which we detail in App.\ \ref{app:maj}. We note that a recent work \cite{chinzei2024} has made significant progress in this direction, showing that the commuting-block generator construction offers and optimal trade-off between expressivity and gradient efficiency. A further work \cite{coyle2024} increases expressivity by using probabilistic mixtures of commuting circuits.

%Could there be a construction that leads to a full-dimensional dynamical Lie algebra, thus allowing for universal quantum computation within the parameterised part of the ansatz? 

An alternative route to increased expressivity would be to consider greedy layer-wise training. Here, one would train a sequence of layers in the order they are performed, where each layer contains mutually commuting generators. By viewing previously trained layers as part of the initial unitary $V$ in a commuting-generator circuit, parameters can be trained in parallel. Note this goes beyond the commuting-block circuit ansatz since we no longer require a fixed commutation relation between generators in differing layers. Although this strategy allows for universal quantum computation \cite{lloyd2018quantum}, it remains to be seen if layer-wise training is sufficient for good performance at scale. 

Our results could also have significant impact across fields other than machine learning, in which quantum circuit optimisation plays a role. For example, can one construct problem-inspired ans\"{a}tze based on our circuits in a similar spirit to QAOA? For the $X$-generator ansatz with the choice $V=\mathbb{I}$, it is known that no superpolynomial advantage with respect to the approximation ratio is possible relative to classical algorithms \cite{lee2021progress}. However, considering other choices of $V$ or non-commuting layers via the commuting-block class may lead to faster variational algorithms for optimisation. 

One other area that is in critical need of resource reductions is quantum chemistry, where the need to evaluate a large number of terms that comprise the Hamiltonian can render optimisation unfeasible in practice \cite{vqe_review}. Much effort has been put into reducing this burden e.g.~via clever groupings of mutually anticommuting terms \cite{Verteletskyi_Izmaylov_20,Yen_Izmaylov_20} or additional preprocessing steps of $\mathcal{H}$ \cite{Huggins_Babbush_21}, and our work can be seen as an additional method that attempts to find savings by exploiting commutation between circuit generators. It is nevertheless unclear whether our class of circuits can provide a useful structure for variational quantum chemistry: for example, parameterised gate sets that respect particle-number conservation in the Jordan-Wigner form do not form a commuting set and cannot obviously be put into a commuting-block form. 

Finally, a key question with regards to quantum machine learning is to identify interesting symmetries that can be encoded into our circuits beyond that of the 1D translation symmetry considered here. Here, there is a possible link to quantum error correcting codes. Namely, suppose we have a data symmetry whose representation on the data-encoded quantum states corresponds to the logical operators of some error-correcting code. Then, since the code stabilisers commute with these operators, they can be used as generators to construct equivariant gates. Identifying data encodings and data symmetries that fit this framework may therefore lead to specific applications that leverage the backpropagation scaling of the quantum circuits proposed in this work.

\section{Outlook}
The overriding message we hope to convey with this work is that there is a strong need to move away from generic, unstructured models in quantum machine learning and towards specific circuit structures that are engineered to have desirable features. Indeed, works that consider unstructured quantum circuits as viable machine learning models often end up painting an overly pessimistic picture due to the disadvantages of working with an unreasonably large model class \cite{mcclean2018barren,anschuetz2022quantum,anschuetz2021critical,campos2021}. With regards to gradient estimation, more focus on structured, non-universal circuits like those presented in this work may be necessary, since backpropagation scaling may be impossible for general models \cite{abbas2023quantum}.
 
% Further progress with respect to the training efficiencies of quantum models is crucial, since the intense computational effort needed to train from data means that any scaling that is greater than linear in the number of data points or model parameters often means training is intractable in practice. 

Although our work presents a potential route to efficient training for supervised problems such as classification or regression, we stress that more progress may be needed for generative problems, where access to log-probabilities---often required to train large classical generative models---is uncomfortably absent in quantum machine learning. In the long run, we hope that more focus on finding circuit architectures that can be trained at scale will lead to a `neural network' moment in quantum machine learning, marked by the emergence of the specific building blocks necessary to construct scalable and powerful quantum learning models. 

\section{Acknowledgements}
JB is grateful to Patrick Huembeli for initial discussions about this project. All authors thank Maria Schuld, Nathan Killoran and Josh Izaac for useful input and comments.

\bibliographystyle{quantum}
\bibliography{refs}

\appendix

\section{Proof of Corollary \ref{thm:sim}}\label{app:simproof}
The theorem follows from the results of \cite{ni2012commuting} where it is shown that it is possible to estimate the expectation value 
\begin{align}\label{comsim}
    \bra{x}U^{\dagger}(Z\otimes\mathbb{I}\otimes\cdots\otimes\mathbb{I})U\ket{x}
\end{align}
for any circuit $U$ that is comprised of exponentiated commuting Pauli operators $e^{-i\theta P}$  to precision $\epsilon$ in poly($N,\frac{1}{\epsilon}$) runtime. Here $\ket{x}$ is any computational basis input. For a circuit with $\mathcal{H}=F^\dagger Z_1 F$ we need to estimate
\begin{align}
    C(\vec{\theta})&=\bra{0}V^{\dagger} U^{\dagger}(\vec{\theta})F^{\dagger} Z_1 F {U}(\vec{\theta})V\ket{0}\\ &= \bra{0}V^{\dagger}F^{\dagger} \tilde{U}^{\dagger}(\vec{\theta}) Z_1 \tilde{U}(\vec{\theta})FV\ket{0}\label{killer}
\end{align}
for some other commuting circuit $\tilde{U}(\vec{\theta})$ with Pauli generators $
F G_j F^\dagger$. If $FV\ket{0}=\ket{x}$ our expression therefore takes the same form as \eqref{comsim}. Since sampling the quantum circuit results in an $\epsilon$-estimate of $C(\vec{\theta})$ in run time $\frac{1}{\epsilon^2}$, it follows that the same precision can be achieved by the classical algorithm with at most polynomial overhead in $N$. 

We note that for `strong' classical simulation of expectation values (i.e.\ exponential accuracy), classical simulability does not hold in general \cite{ni2012commuting}, however the quantum circuit itself cannot achieve this precision.

\begin{figure}
    \centering
    \includegraphics[width=\columnwidth]{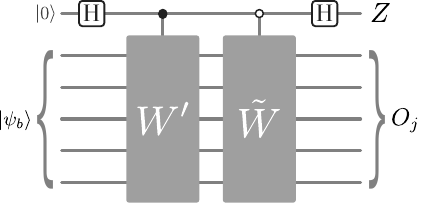}
    \caption{The circuit  used to estimate the partial derivatives of a commuting block.}
    \label{fig:lcu}
\end{figure}

% On the bright side, this observation does imply that IPQ circuits can be efficiently optimised classically. If the ultimate goal is to sample from a distribution (e.g. QAOA) which is provably hard, this could be a useful method to avoid expensive optimisation on a QPU. 

\section{Gradient estimation of commuting-block circuits}\label{app:blockgrads}
Here we prove Thm.\ \ref{thm:blocks}. Consider a commuting-block ansatz with $B$ blocks. The cost $C(\vec{\theta})$ can be written as
\begin{align}
    C(\vec{\theta}) = \bra{\psi_b}W^\dagger \mathcal{H} W \ket{\psi_b}
\end{align}
where 
\begin{align}
    \ket{\psi_b}=U_{b}(\vec{\theta}^{b})\cdots U_2(\vec{\theta}^2)U_1(\vec{\theta}^1) V\ket{0}
\end{align}
is the state after the $b^{\text{th}}$ block and 
\begin{align}\label{wdef}
    W = U_{B}(\vec{\theta}^{B})\cdots U_{b+2}(\vec{\theta}^{b+2})U_{b+1}(\vec{\theta}^{b+1})
\end{align}
is the parameterised circuit after the $b^{\text{th}}$ block. The partial derivatives of the $b^{\text{th}}$ block are then 
\begin{align}\label{pdevblock1}
\frac{\partial C}{\partial \theta^b_j} = \bra{\psi_b}(iG_jW^\dagger \mathcal{H} W -iW^\dagger \mathcal{H}WG_j)\ket{\psi_b},
\end{align}
where $G_j$ is the generator of the $j^{\text{th}}$ unitary in block $b$. Since all generators in the block share the same commutation relation with other blocks, we have 
\begin{align}
    WG_j = G_j\tilde{W} \quad \forall\;G_j
\end{align}
for some other unitary $\tilde{W}$. For example, if generators from all other blocks anticommute with those of block $j$ then
$\tilde{W}(\vec{\theta})=W(-\vec{\theta})$
since
\begin{align}
    G_je^{i\theta_k G_k} = e^{-i\theta_k G_k}G_j
\end{align}
for $\{G_k,G_j\}=0$. We may now write \eqref{pdevblock1} as 
\begin{align}\label{pdevblock2}
\frac{\partial C}{\partial \theta^b_j} 
= \bra{\psi_b}(\tilde{W}^\dagger iG_j\mathcal{H} W -(-1)^{g_j} W^\dagger iG_j\mathcal{H}\tilde{W})\ket{\psi_b},
\end{align}
Where $g_j = 0$ or $g_j=1$, depending on whether $G_j$ commutes or anticommutes with $\mathcal{H}$.
Defining $O_j=i^{g_j}G_j\mathcal{H}$, \eqref{pdevblock2} takes the form 
\begin{align}\label{pdevblock3}
\frac{\partial C}{\partial \theta^b_j} 
&= \bra{\psi_b}\de{\tilde{W}^\dagger O_j W' + (W')^\dagger O_j\tilde{W}}\ket{\psi_b}\\
&=\frac12\left[\bra{\psi_b}\de{\tilde{W}^\dagger + (W')^\dagger} O_j \de{W'+\tilde{W}}\ket{\psi_b}\right.\nonumber\\
&\left.+\bra{\psi_b}\de{\tilde{W}^\dagger - (W')^\dagger} O_j \de{W'-\tilde{W}}\ket{\psi_b}\right]\\
&=\frac12\left[\braket{O_j}_{L_W^+\ket{\psi_b}}-\braket{O_j}_{L_W^-\ket{\psi_b}}\right]
\end{align}
where we first defined the modified unitary $W'=i^{1-g_j}W$ and then introduced the linear combinations of unitaries $L_W^\pm=\tilde{W}\pm W'$.

Consider the circuit depicted in Fig.\ \ref{fig:lcu}, which applies controlled-$W'$ instead of $W$ in the original circuit, followed by a controlled-$\tilde{W}$ controlled on the first qubit being in state $\ket{0}$. The prepared state is
\begin{align}
    \ket{\phi_b} 
    &= \frac12\De{\ket{0}L^+_W\ket{\psi_b}+\ket{1}L^-_W\ket{\psi_b}}.
\end{align}
If we then measure the expectation value of $\tilde{O}_j=2(Z\otimes O_j)$ on this state we obtain
\begin{align}
    \braket{\tilde{O}_j}_{\ket{\phi_b}}
    =& \frac12 \left[\bra{0}Z\ket{0}\bra{\psi_b}(L^+_W)^\dagger O_j L^+_W\ket{\psi_b}\right. \nonumber\\
    &\left.+\bra{1}Z\ket{1}\bra{\psi_b}(L^-_W)^\dagger O_j L^-_W\ket{\psi_b}\right]\nonumber\\
    =& \frac{\partial C}{\partial \theta^b_j}.
\end{align}
Similar to the single-block case we may confirm that
\begin{align}
    [O_j,O_k] = 0\quad \forall \;j,k. 
\end{align}
for generators with the same value of $g_j$. The observables can therefore be simultaneously diagonalised and evaluated in parallel and we need to evaluate at most $2$ circuits to estimate all partial derivatives of any given block, one for the generators that commute with $\mathcal{H}$ and one for those that anticommute. Since the derivative is estimated as a standard circuit expectation value, its variance scales as $1/M$ with $M$ the number of shots.   

For the last block we note that $\tilde{W}=W=\mathbb{I}$ so that we may use parallel estimation without any auxiliary qubits. Alternatively, this case can be interpreted as single-block commuting-generator circuit by absorbing all other blocks into $V$. The total number of circuits required for full gradient estimation of $B$ blocks is therefore $2B-1$.

\section{Maximal set of mutually anticommuting Pauli tensors}\label{app:maj}
A maximal set of $2N+1$ mutually anticommuting Pauli products can be found as follows. Here we give the construction for five qubits; the generalisation to $N$ qubits is straightforward. In the Jordan-Wigner representation, the Majorana operators look like:
\begin{align}
\begin{matrix}
    P_1 = & X&I&I&I&I \\
    P_2 = &Z&X&I&I&I \\
    P_3 = &Z&Z&X&I&I \\
    P_4 = &Z&Z&Z&X&I \\
    P_5 = &Z&Z&Z&Z&X \\
    P_6 = &Y&I&I&I&I \\
    P_7 = &Z&Y&I&I&I \\
    P_8 = &Z&Z&Y&I&I \\
    P_9 = &Z&Z&Z&Y&I \\
    P_{10} = &Z&Z&Z&Z&Y,
\end{matrix}
\end{align}
which can be seen to mutually anticommute. To these we may add $P_{11}=ZZZZZ$, giving the desired set of $2N+1=11$ mutually anticommuting operators. 

\section{The diagonalising unitaries for the \texorpdfstring{$X$}{X}-generator ansatz}\label{app:xgrads}
Here we show that the circuits described in the main text diagonalise the operators $O_j=2iG_j\mathcal{H}$ and $O_{jk}=-4G_jG_k\mathcal{H}$ in the product $X$ basis for $\mathcal{H}$ given by \eqref{zobs}. That is, we show
\begin{align}\label{appdiag1}
    D O_j D^\dagger = 2G_j
\end{align}
and 
\begin{align}\label{appdiag2}
    D H_{zy}^{(1)} O_{jk} H_{zy}^{(1)} D^\dagger = (-1)^{\beta_1}4X^{(1)}G_jG_k.
\end{align}
We first prove \eqref{appdiag1}. Let us denote the number of $X$ operators of $G_j$ on the first $m$ qubits as $m_j$. Since we need only consider generators that anticommute with $\mathcal{H}$, we have that $m_j$ is odd and the $O_j$ take the form 
\begin{align}\label{Oj1}
    O_j = 2s_j P_{j,1}\otimes \cdots \otimes  P_{j, m} \otimes G_{j, m+1}\otimes \cdots \otimes G_{j, N},
\end{align}
where $P_{j, r}\in{Z,Y}$ with an odd number of $Y$s. The sign factor takes the form
\begin{align}\label{factor}
        s_j
        = i^{m_j-1}
        =\begin{cases}
            1 & \text{ for } m_j \nospacemod 4 = 1\\
            -1 &  \text{ for } m_j \nospacemod 4 = 3
        \end{cases}.
\end{align}
Since the $O_j$ all contain $X$ or $\mathbb{I}$ operators on the last $n-m$ qubits, $D$ needs only to act on the first $m$ qubits. Our task is therefore to diagonalise operators $\{P^j_1\otimes P^j_2\otimes\cdots\otimes P^j_m\}$ where $P^j_i\in\{Y,Z\}$ and there are an odd number of $Y$'s. 

\begin{figure}
    \centering
\includegraphics[width=0.9\columnwidth]{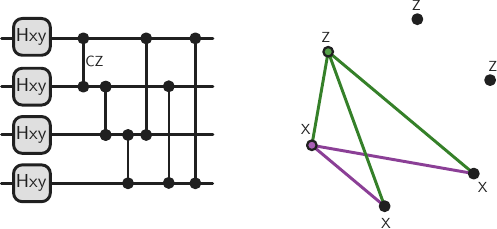}
    \caption{The diagonalising unitary (left) and a visual aid for the proof of App.\ \ref{app:xgrads} (right). Note that the order of operations in the diagonalising circuit matches the order in which we apply them to the original (non-diagonal) observable, because we implement $D^\dagger$ instead of $D$ and because the operations are applied to the observable in reversed order compared to the quantum state.}
    \label{fig:diagcircuit}
\end{figure}

To do this we first apply an X-Y Hadamard to each qubit, so we now have operators 
\begin{align}
    (-1)^{m-m_j}P_{j,1}\otimes P_{j,2}\otimes\cdots\otimes P_{j,m}
\end{align}
with $P_{j,i}\in\{Z,X\}$ with an odd number of $X$'s. We then apply the sequence of CZ operators between every pair of qubits. To make things clearer, we consider a specific operator which we show in Fig.\ \ref{fig:diagcircuit}. We call the qubits that have a $Z$/$X$ operator `Z qubits' or `X qubits'. Since $CZ(Z\otimes Z)CZ = Z\otimes Z$, CZ operators that act of $Z$ sites have no effect. Those that act between a given $Z$ site and an $X$ site (green lines in Fig.\ \ref{fig:diagcircuit}) have the effect of changing the $Z$ site to an identity operator. This follows from the property
\begin{figure*}

    \centering
    \includegraphics[width=\textwidth]{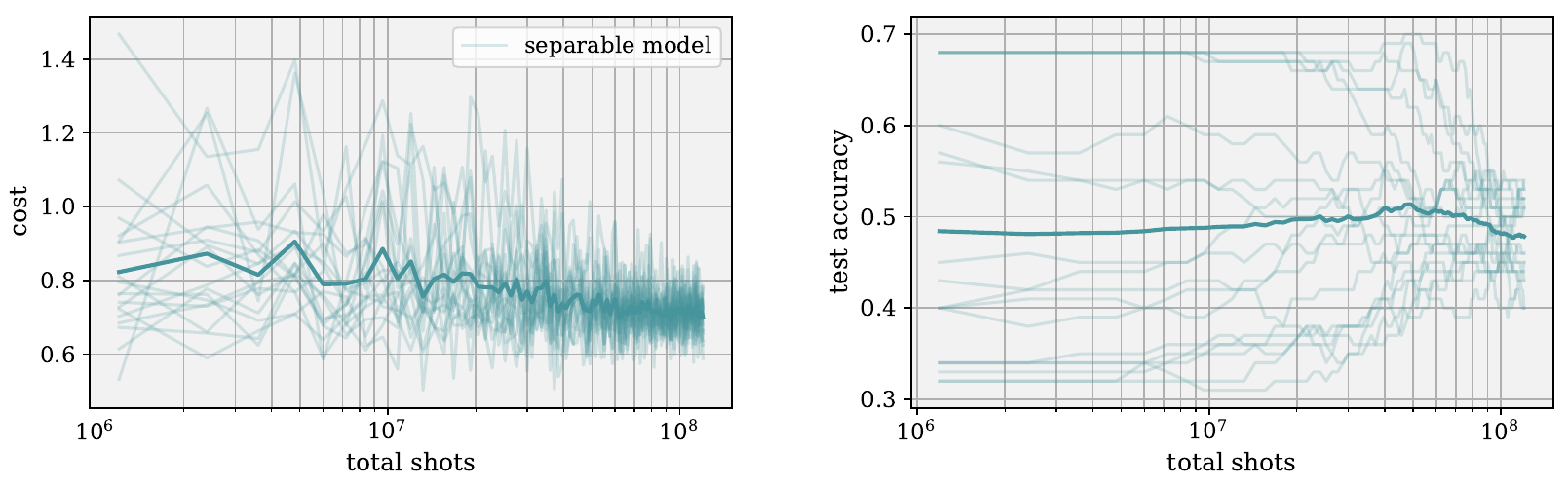}
    \caption{Results for the 16-qubit benchmarking problem of Sec.\ \ref{sec:numerics} using the separable model (Model D).}
    \label{fig:modelD}
\end{figure*}
\begin{align}\label{conj}
    CZ(X\otimes \mathbb{I})CZ = X\otimes Z.
\end{align}
Since there are an odd number of $X$ operators, the combined effect is to multiply the $Z$ site by an odd number of $Z$ operators, thus converting it to an identity. We are thus left with all the CZ operators that act between two $X$ sites (purple lines). From \eqref{conj} and $XY=iZ$ we have  
\begin{align}\label{conj2}
    CZ(X\otimes X)CZ = Y\otimes Y, \\ CZ(Y\otimes X)CZ = - X\otimes Y. 
\end{align}
Let us label the $X$ sites as $1,\cdots,m_j$. From \eqref{conj2}, if we apply all CZ operators between the first site and the rest, the effect of this is (i) alternate the operator at site 1 between $X$ and $Y$ an even number of times, thus ending as an $X$ operator, and (ii) pick up a factor of $(-1)^{\lfloor(m_j-1)/2\rfloor}=-1$. In this process the remaining $X$ sites now have a $Y$ operator. Applying all CZs between site 2 and the rest now results in an odd number of flips (leaving site 2 with an $X$ operator), and picking up a factor $(-1)^{\lfloor(m_j-2)/2\rfloor}=+1$. 
The remaining $m_j-2$ sites now have an $X$ operator as before. Continuing this process we see that the $X$ operators remain unchanged, but we pick up a factor
 \begin{align}\label{factorap}
        (-1)^{\left(\frac{m_j-1}{2}\right)^2}
        =\begin{cases}
            1 & \text{ for } m_j \nospacemod 4 = 1\\
            -1 &  \text{ for } m_j \nospacemod 4 = 3
        \end{cases}.
\end{align}
This factor matches precisely the factor $s_j$ in \eqref{factor} which multiplies the operators. It follows that the action of the whole circuit $D$ is given by \eqref{appdiag1}. 

We now prove \eqref{appdiag2}. The possible $O_{jk}$ take a similar form to \eqref{Oj1}:
\begin{align}\nonumber
    O_{jk} = 4s_{jk} P_{j, 1}\otimes \cdots \otimes  P_{j, m} \otimes G_{j, m+1}\otimes \cdots \otimes G_{j, N}, 
\end{align}
where now we have an even number $m_{jk}$ of $Y$ operators on the first $m$ qubits and 
\begin{align}\label{factor2}
        s_{jk}
        =(-1)^{\frac{m_{jk}}{2}+1}
        =\begin{cases}
            1 & \text{ for } m_{jk}\nospacemod 4 = 2\\
            -1 &  \text{ for } m_{jk}\nospacemod 4 = 0
        \end{cases}.
\end{align} 
Note that if we apply an z-y Hadamard operator on the first qubit, we map any $O_{jk}$ to an operator that now has an odd number of $Y$'s on the first $m$ qubits, which is precisely the case we had before. After applying $D$, from \eqref{factorap} and \eqref{factor2} we see that the sign factor we obtain matches $s_{jk}$ if $G_{j}G_k$ has an $X$ operator on the first qubit; if not we pick up a factor $-1$. We therefore find $D H_{zy}^{(1)} O_{jk} H_{zy}^{(1)} D^\dagger = (-1)^{\beta_1}4X^{(1)}G_jG_k$ as claimed.

\section{finite-shot numerics}\label{app:finiteshot}
\begin{figure}
    \centering
    \includegraphics[width=\columnwidth]{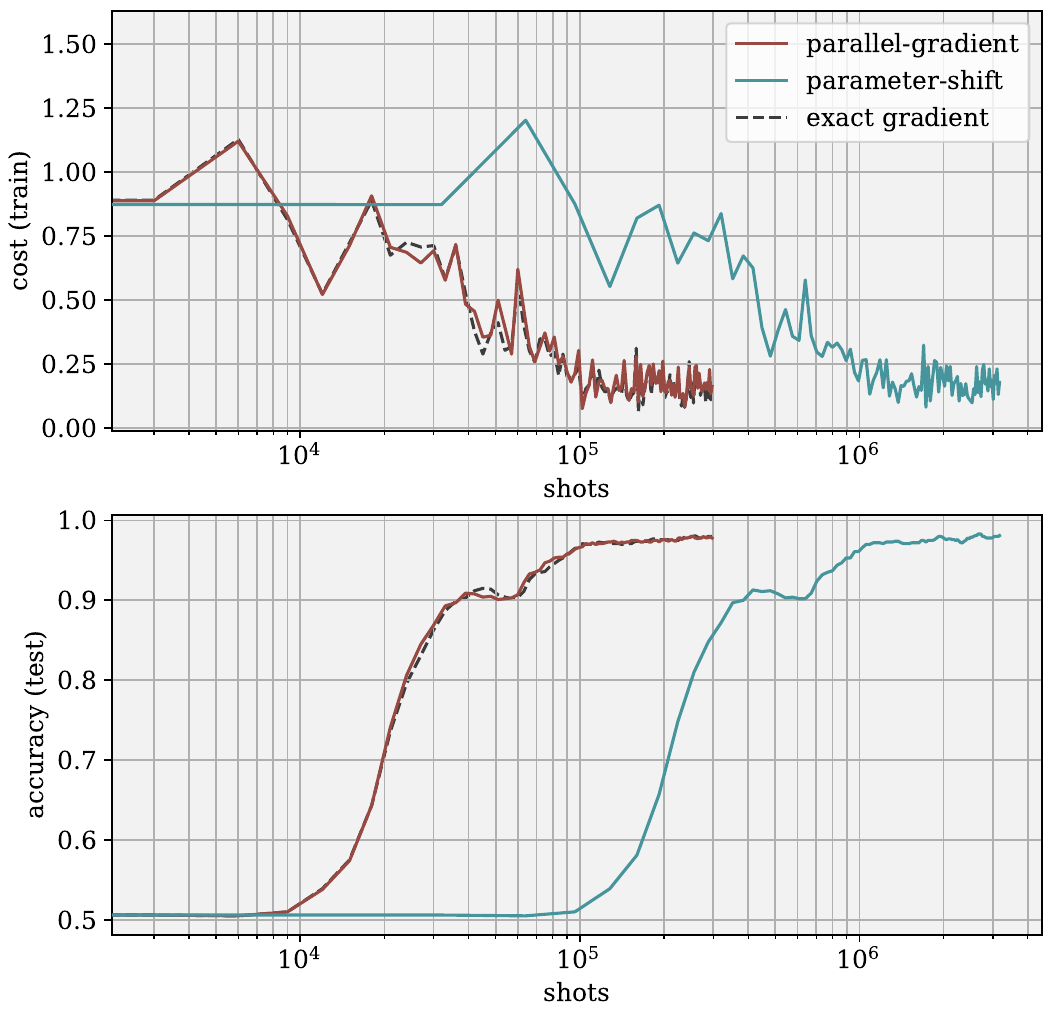}
    \caption{Results for the 6-qubit problem with finite shots described in App.\ \ref{app:finiteshot}, using 500 shots per circuit. }
    \label{fig:finiteshot}
\end{figure}

In Fig.\ \ref{fig:finiteshot} we show training and accuracy curves for Model A from Sec.\ \ref{sec:numerics} for a 6-dimensional bars and dots problem, using finite-shot analysis and 500 shots per quantum circuit used in the gradient evaluation. Here, the model has access to up to 6-body generators, i.e. $K=6$. The black dotted curve corresponds to the learning trajectory if the exact gradient were provided instead of that given by the finite shots. As can be seen, the finite shot updates approximate well the exact behaviour. 

\section{Details on the required shot counts}\label{app:shot_estimation}
Here we present a few details on the numbers of shots required to compute a gradient for the four models discussed in Sec.\ \ref{sec:numerics}.

For Model A, the calculation of the required shots given in the main text is already sufficiently precise and we have $M_{A}=M\cdot d$.

For Model B, the number of Pauli generators can be evaluated to be
\begin{align}
    \mathcal{P} = L\left(2d+\frac{d^2-d}{2}\right)=L\cdot \frac{d}{2} \cdot (d+3),
\end{align}
where the parentheses in the first expression consists of the contributions for the two single-qubit layers and for the block of two-qubit gates.
Each generator is a Pauli word and thus requires two parameter-shifted circuits, leading to a na\"ive count of $M_B=MLd(d+3)$.
However, in the last layer of the circuit we may exploit commutativity within the block of two-qubit gates in order to execute the parameter-shift rule for $d/2$ of these gates simultaneously. This reduces the number of different circuits for the last two-body block from $d^2-d$ to $\frac{d^2-d}{d/2}=2(d-1)$. The resulting corrected shot count for a full gradient evaluation of Model B is then
\begin{align}
    M_B
    &=M\left[(L-1)d(d+3)+(6d-2)\right]\nonumber\\
    &=M \left[d^2 (L-1) +3d(L+1)-2\right].
\end{align}

In Model C, we use $-1+\sum_{j=1}^{\log_2 d} 2^j = 2d-3$ two-qubit convolution blocks with $10$ gates each, where we assumed that $d$ is a power of two. 
Two of these gates are controlled Pauli rotations, which require four instead of two parameter-shifted circuits each.
Thus we require $(2d-3) \cdot (8\cdot 2+2\cdot 4)=24(2d-3)$ circuits for the gradient w.r.t.\ the convolution parameters.
In addition, the model uses $\sum_{j=0}^{\log_2 d-1}2^j = d-1$ pooling blocks, with two controlled Pauli rotations each.
This leads to $(d-1)\cdot (2\cdot 4)=8(d-1)$ circuits to compute the gradient w.r.t.\ the pooling parameters.
Overall, we find the shot cost for Model C to be
\begin{align}
    M_C = M\left[24(2d-3)+8(d-1)\right]
    =M(56d-80).
\end{align}

Lastly, the gradient for the separable Model D can be computed in parallel between the different qubits, as they do not interact at all.
There are $3d$ Pauli generators, of which $d$ can be treated in parallel, leading to shot cost of $M_D=M\cdot 2\cdot 3d/d=6M$.

\end{document}